\documentclass[secnumarabic,aps,pra,amsfonts,twocolumn,showkeys,floatfix]{revtex4}

\usepackage{bm,longtable,ams,amsfonts}
\usepackage{placeins}
\usepackage{graphicx}
\newcommand{\gl}[1]{(\ref{#1})}

\begin{document}

\title{Theoretical investigation of the magnetic structure in
  YBa$_2$Cu$_3$O$_6$}  
\author{Ekkehard Kr\"uger}
\affiliation{Max-Planck-Institut f\"ur Metallforschung, D-70506 Stuttgart,
  Germany}
%
\date{5. August 2006}
\begin{abstract}
  As experimentally well established, YBa$_2$Cu$_3$O$_6$ is an
  antiferromagnet with the magnetic moments lying on the Cu sites. Starting
  from this experimental result and the assumption, that nearest-neighbor
  Cu atoms within a layer have exactly antiparallel magnetic moments, the
  orientation of the magnetic moments has been determined within a
  nonadiabatic extension of the Heisenberg model of magnetism, called
  nonadiabatic Heisenberg model. Within this group-theoretical model there
  exist four stable magnetic structures in YBa$_2$Cu$_3$O$_6$, two of them
  are obviously identical with the high- and low-temperature structure
  established experimentally. However, not all the magnetic moments which
  appear to be antiparallel in neutron-scattering experiments are exactly
  antiparallel within this group-theoretical model. Furthermore, within
  this model the magnetic moments are not exactly perpendicular to the
  orthorhombic $c$ axis.
\end{abstract}
\maketitle

\section{Introduction}
The magnetic structure of YBa$_2$Cu$_3$O$_6$ has been studied by
neutron-scattering experiments with high accuracy \cite{shamoto,tranquada}.
YBa$_2$Cu$_3$O$_6$ is an insulating antiferromagnet with the magnetic
moments lying on the Cu sites. Two magnetic structures have been observed:
a high-temperature structure below about 410 K and a low-temperature
structure below 15 K.

In the framework a nonadiabatic extension of the Heisenberg model
\cite{hei}, called nonadiabatic Heisenberg model (NHM) \cite{enhm}, a
theoretical investigation of all the stable magnetic structures in
YBa$_2$Cu$_3$O$_6$ is possible. Starting from the experimental result that
the magnetic moments lie on the Cu sites and that nearest-neighbor Cu atoms
within a layer have exactly antiparallel magnetic moments, the NHM gives a
clear description of the orientation of the magnetic moments.

Within the NHM, the electrons in narrow, half-filled bands may lower their
Coulomb energy by occupying an atomiclike state as defined by Mott
\cite{mott} and Hubbard \cite{hubbard}: as long as possible the electrons
occupy localized states and perform their band motion by hopping from one
atom to another. The localized states are represented by localized
functions depending on an additional coordinate\ $\vec q$~related to the
nonadiabatic motion of the center of mass of the localized state.  The
introduction of the new coordinate allows a consistent specification of the
atomiclike motion having a {\em lower energy} than a purely bandlike
motion. The nonadiabatic localized functions {\em must} be adapted to the
symmetry of the crystal so that the nonadiabatic Hamiltonian $H^n$ has the
correct commutation properties.

The nonadiabatic localized functions may be approximated by the best
localized Wannier functions when the nonadiabatic motion of the centers of
mass again is disregarded. These ``adiabatic'' and the nonadiabatic
functions will not differ so strongly that their symmetry is altered at the
transition from the adiabatic to the nonadiabatic system.  Hence, suitable
nonadiabatic localized states allowing an atomiclike motion of the
electrons exist if and only if best localized Wannier functions exist which
are adapted to the symmetry of the crystal.

The NHM is a purely group-theoretical model which makes use of the magnetic
group of the spin structure and the symmetry of the Bloch functions of the
narrowest partly filled bands in the band structure of the considered
material. It does not specify the physical nature of the exchange mechanism
generating the magnetic structure but makes clear statements about all the
stable magnetic structures of this material.

Within the NHM, it cannot be distinguished between orbital and spin
moments.  Therefore, in this paper I always speak of ``magnetic moments''
which may consist of both orbital and spin moments.

\section{Magnetic bands}
\begin{figure*}[!]
  \includegraphics[width=.9\textwidth,angle=0]{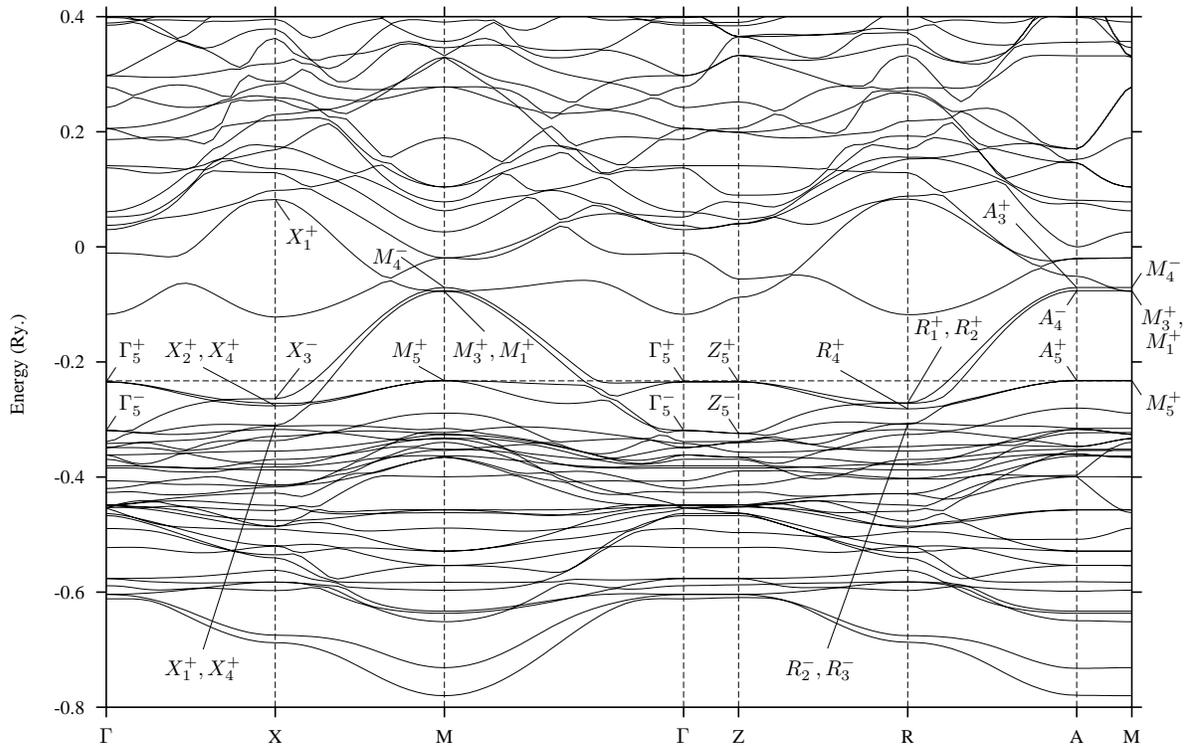}
  \caption{
    Band structure of tetragonal YBa$_2$Cu$_3$O$_6$ as calculated by Ove
    Jepsen \cite{jepsen}, with symmetry labels determined by the author.
  \label{fig:bandstr}
  }
  \end{figure*}

The Bloch functions of the narrowest conduction bands of paramagnetic
YBa$_2$Cu$_3$O$_6$ are labeled by the representations
\begin{equation}
\label{eq:1}
\begin{array}{l}
\Gamma^+_5, \Gamma^-_5;\\  X^+_2, X^+_4, X^-_3, X^+_1, X^+_4;\\ 
Z^+_5, Z^-_5; \\ R^+_1, R^+_2, R^+_4, R^-_2, R^-_3 ;\\ 
A^+_3, A^-_4, A^+_5; \\ M^-_4, M^+_3, M^+_1, M^+_5,
\end{array}
\end{equation}
see Fig.~\ref{fig:bandstr}.  


Table~\ref{tab:10} lists all the bands in tetragonal YBa$_2$Cu$_3$O$_6$
with symmetry-adapted and best localized Wannier functions \cite{enhm}
situated on the Cu sites. Strictly speaking, each band is a ``band
complex'' consisting of three single bands or ``branches'' because there
are three Cu atoms in the unit cell.

None of the bands in Table~\ref{tab:10} contains the representation
$\Gamma^+_5$ or $\Gamma^-_5$. This observation alone shows that it is not
possible to represent the narrowest partly filled bands of tetragonal
YBa$_2$Cu$_3$O$_6$ by best localized Wannier functions which are situated
on the Cu sites and adapted to the symmetry of the tetragonal crystal.
Consequently, in tetragonal YBa$_2$Cu$_3$O$_6$ the conduction electrons
{\em cannot}\ occupy the energetically favorable atomiclike state because
suitable localized states do not exist.

In magnetic materials the electrons have another way to occupy an
atomiclike state. In any magnetic structure, the symmetry of the
paramagnetic system is lowered. The construction of best localized Wannier
functions adapted to the {\em reduced} symmetry of the magnetic state may
be possible even if Wannier functions adapted to the symmetry of the
paramagnetic phase do not exist. 

Assume that a given material possesses a narrow, partly filled ``magnetic
band'' related to a special magnetic structure $S$. That means that the
Bloch functions of this energy band can be unitarily transformed into best
localized Wannier functions which are adapted to the reduced symmetry of
the structure $S$. Assume further that the structure $S$ actually exists in
this material. Then the electrons may lower their Coulomb energy by
occupying an atomiclike state since now localized states with the correct
symmetry exist.

This idea may be interpreted as follows~\cite{ea,ef,enhm}: {\em If} a
paramagnetic material possesses a narrow, partly filled magnetic band
related to a magnetic structure $S$, {\em then} the electron system {\em
  activates} a spin dependent exchange mechanism producing this structure.
Such a process is allowed within the NHM because this model goes beyond the
adiabatic approximation. In the adiabatic system the localized electrons
move on {\em rigid}~orbitals (represented by the best localized Wannier
functions) in the average potential of the other electrons.  Within the
NHM, on the other hand, the localized states are represented by localized
functions depending on the above mentioned additional coordinate $\vec q$
related to the nonadiabatic motion of the center of mass of the localized
states.  The localized electrons now move on temporarily modified orbitals
(each characterized by a definite value of~$\vec q$\,) in a potential
depending on which of the adjacent localized states are occupied at
present, and on the present positions of the electrons in these states.
The nonadiabatic orbitals are still symmetry-adapted on the average of
time, but not at any moment. The formation of the magnetic structure $S$
enables the electrons to occupy the energetically favorable atomiclike
state.  Hence, the correlated nonadiabatic motion of the electrons occurs
in such a way that the exchange energy $\Delta E$ (depending crucially on
the exact form of the localized orbitals) is maximum for this structure
$S$.

This interpretation leads to the definition of magnetic bands
which are responsible for stable magnetic states.  Magnetic bands have
already been identified in Cr \cite{ea}, Fe \cite{ef}, and La$_2$CuO$_4$
\cite{la2cuo4}. These bands are evidently essential to the stability of the
magnetic states observed in these materials.

In this paper, I consider antiferromagnetic structures only. An energy band
of YBa$_2$Cu$_3$O$_6$ is called ``antiferromagnetic band with the magnetic
group $M$'' if the Bloch functions of this band can be unitarily
transformed into Wannier functions which are
\begin{itemize}
\item centered on the Cu sites;
\item symmetry-adapted to $M$; and
\item best localized.
\end{itemize}
Each antiferromagnetic band in YBa$_2$Cu$_3$O$_6$ consists of $\mu$
branches with $\mu = 6$ or $\mu = 12$ being the number of Cu atoms in the
unit cell.

Before I shall consider all the antiferromagnetic bands of
YBa$_2$Cu$_3$O$_6$ in Sec.~\ref{sec:afbands}, I determine in the following
Sec.~\ref{sec:strukt} all the possible magnetic structures in this
material.

\section{Determination of the allowed magnetic structures}
\label{sec:strukt}
Neutron-scattering experiments of antiferomagnetism in YBa$_2$Cu$_3$O$_6$
\cite{shamoto,tranquada} suggest that 
\begin{enumerate}
\item the magnetic moments are localized on the Cu sites and
\item nearest-neighbor Cu atoms within a layer have exactly antiparallel
  magnetic moments. 
\end{enumerate}

\begin{figure*}
\begin{center}
\begin{minipage}{.331\textwidth}
\includegraphics[width=\textwidth,angle=0]{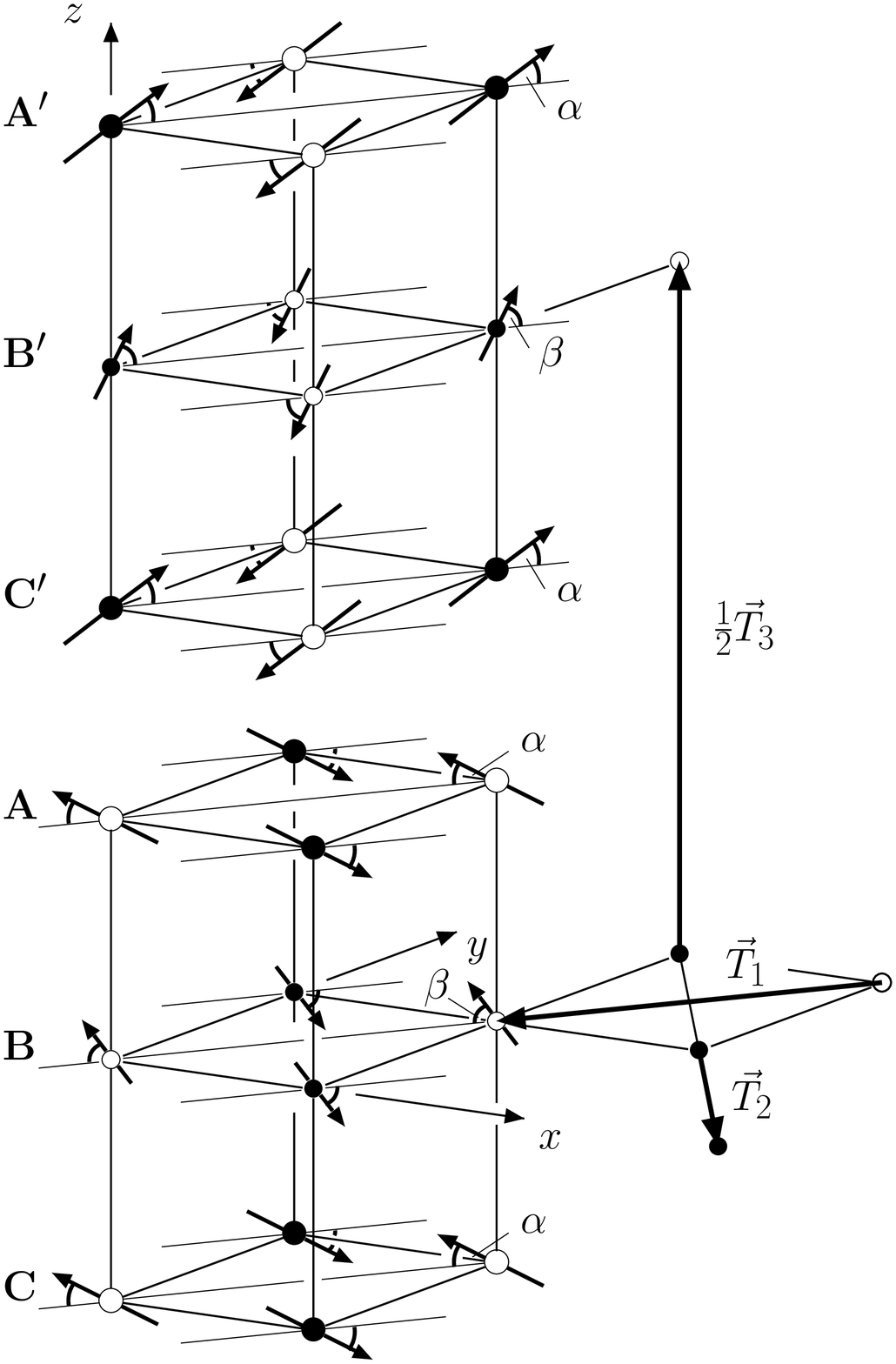}%
\begin{center}
(a)
\end{center}
\end{minipage}\hspace{1.8cm}
\begin{minipage}{.223\textwidth}
\includegraphics[width=\textwidth,angle=0]{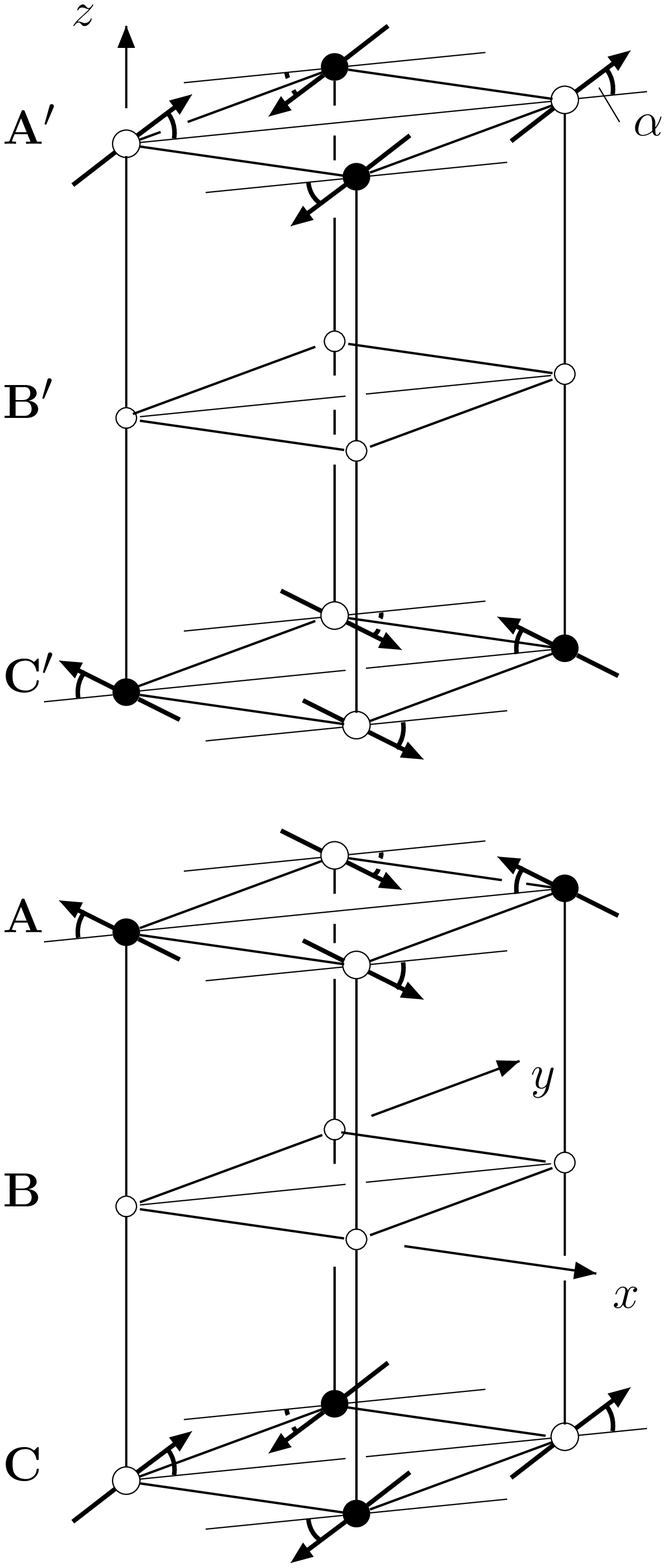}%
\begin{center}
(b)
\end{center}
\end{minipage}
\end{center}
\begin{center}
\begin{minipage}{.331\textwidth}
\includegraphics[width=\textwidth,angle=0]{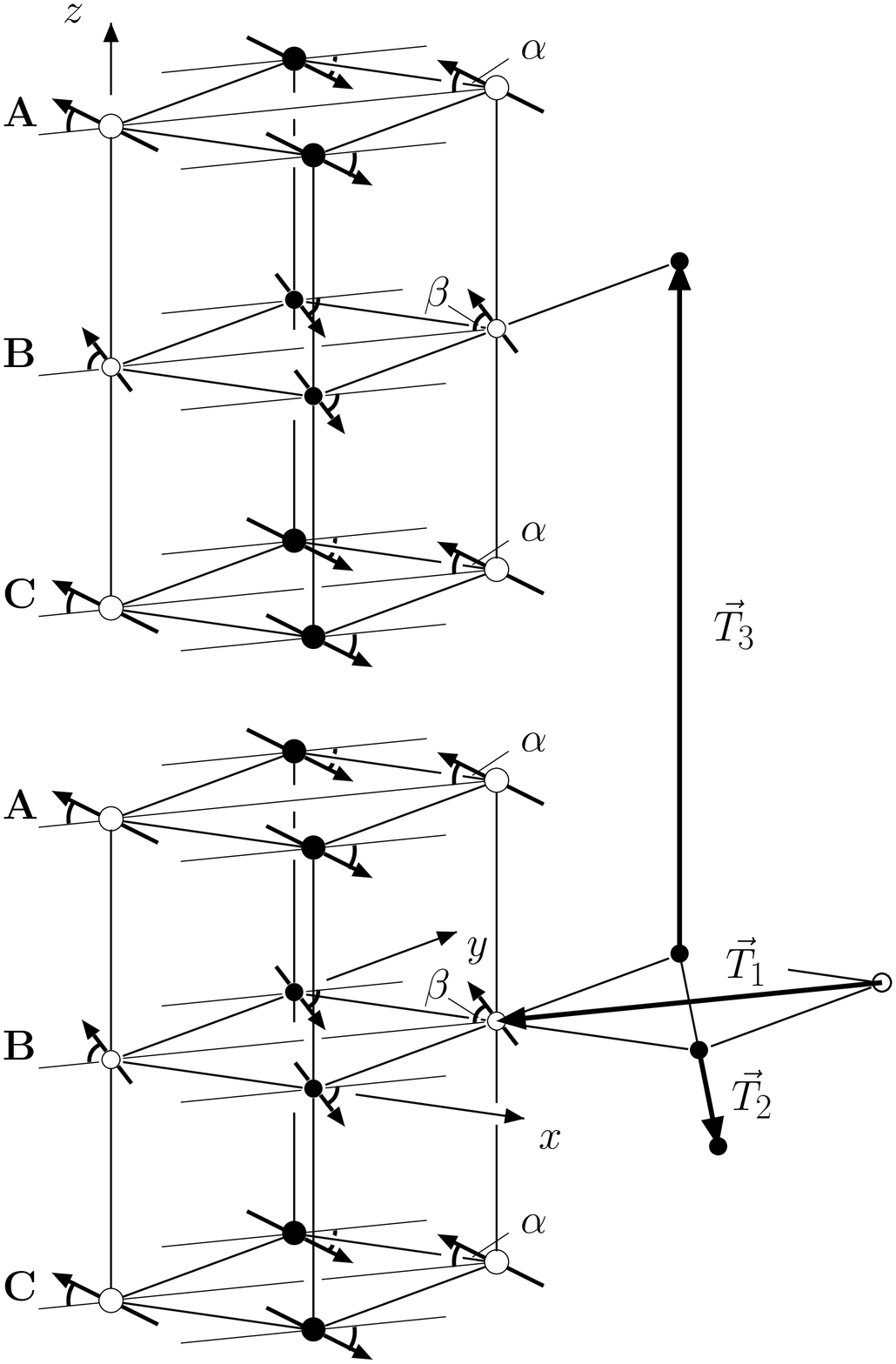}%
\begin{center}
(c)
\end{center}
\end{minipage}\hspace{2cm}
\begin{minipage}{.225\textwidth}
\includegraphics[width=\textwidth,angle=0]{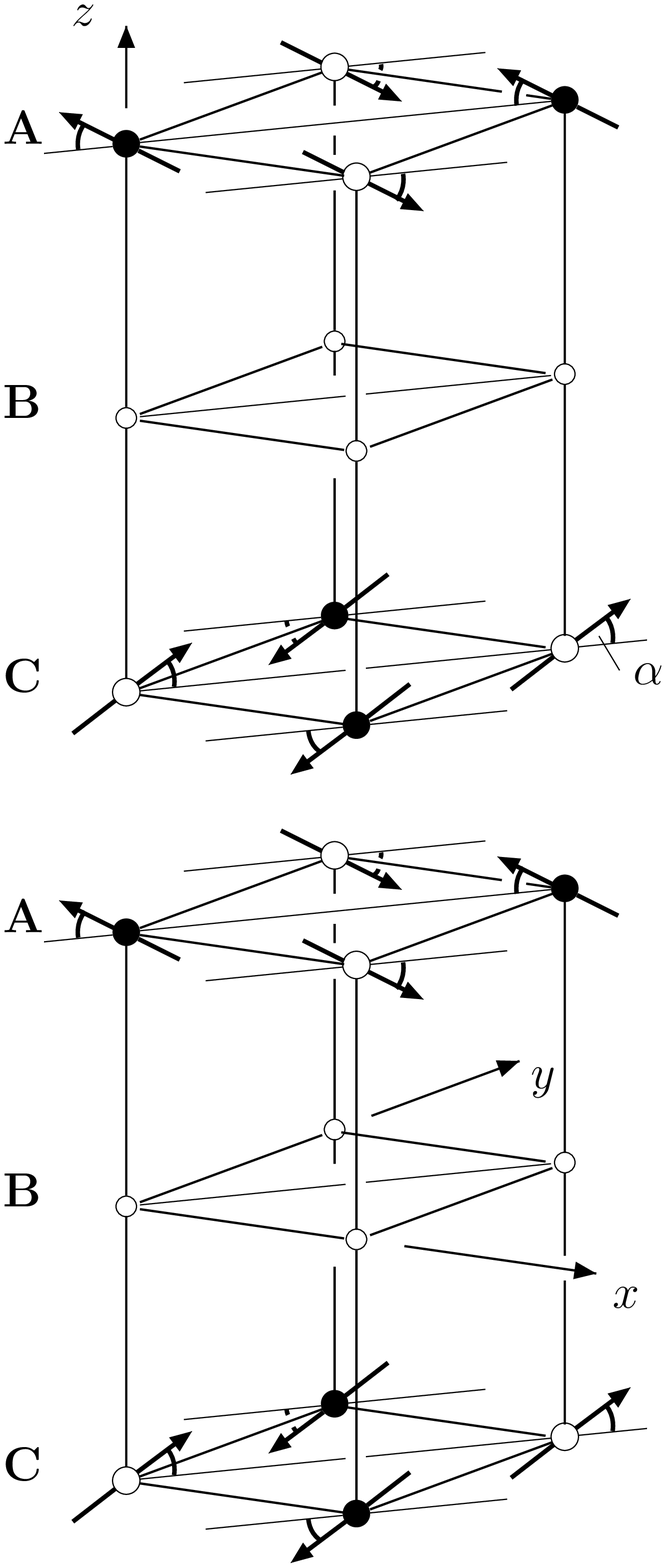}%
\begin{center}
(d)
\end{center}
\end{minipage}\hspace{2cm}
\end{center}
\caption{
  Magnetic structures in YBa$_2$Cu$_3$O$_6$ with the magnetic moments being
  localized on the Cu sites and with nearest-neighbor Cu atoms within a
  layer having antiparallel magnetic moments. The figure shows all the
  structures (a) -- (d) which are stable within the NHM. Solid and open
  circles represent Cu atoms as in Figs.~1 and~3 of
  Refs.~\protect\cite{tranquada} and \protect\cite{shamoto}, respectively,
  to facilitate a comparison with these figures.  The basic translations
  $\vec T_1$, $\vec T_2$, and $\vec T_3$ for the structures (a) and (b) are
  given in
  Fig.~(a), those for (c) and (d) are given in Fig.~(c). 
  The magnetic moments lie exactly in the planes spanned by $\vec T_1$ and
  $\vec T_3$. All the angles $\alpha$ in the $A$, $A'$, $C$, and $C'$
  layers correspond exactly in size and all the angles $\beta$ in the $B$
  and $B'$ layers correspond exactly in size, too. The angles $\alpha$ and
  $\beta$ may be different and are not determined by group theory.
  However, each of the four structures becomes unstable for $\alpha
  \rightarrow 0$ and $\beta \rightarrow 0$ or for $\alpha \rightarrow 90^0$
  and $\beta \rightarrow 90^0$. Thus, the magnetic moments lie neither in
  $\pm \vec T_3$ direction nor perpendicular to $\vec T_3$. The magnetic
  unit cell of the fm-like structure (a) and the af-like structure (b) is
  doubled in $\vec T_3$ direction. In the structures (b) and (d) there are
  no magnetic moments in the $B$ and $B'$ layers. The structures (a), (b),
  and (d) are orthorhombic, (c) is monoclinic. The structures (a) and (d)
  are obviously identical with the experimentally investigated low- and
  high-temperature magnetic structure, respectively, see
  Refs.~\protect\cite{tranquada} and
  \protect\cite{shamoto}. \\
\label{fig:strukt}
}
\end{figure*}

The last point (ii) shows that the magnetic group may be written as
\begin{equation}
  \label{eq:2}
 M = H + \{K|\textstyle\frac{1}{2}\frac{1}{2}0\}H, 
\end{equation}
where $H$ denotes the space group and $K$ stands for the operator of time
inversion. The translational part of the anti-unitary element
$\{K|\frac{1}{2}\frac{1}{2}0\}$ is written in terms of the basic
translations $\vec T_1$, $\vec T_2$, and $\vec T_3$ given in
Fig.~\ref{fig:strukt}.

Within the NHM, not any magnetic group belongs to a {\em stable} magnetic
structure because both the magnetic state $|m\rangle$ and the time-inverted
state
\[
\overline{|m}\rangle = K|m\rangle   
\]
are {\em eigenstates} of the nonadiabatic Hamiltonian $H^n$. Therefore,
a {\em stable} magnetic state $|m\rangle$ complies with two conditions:
\begin{itemize}
\item 
$|m\rangle$ is basis function of a one-dimensional corepresentation
of $M$;
\item $|m\rangle$ and the time-inverted state $\overline{|m}\rangle$ are
  basis functions of a two-dimensional {\em irreducible}\/ corepresentation
  of the gray magnetic group
\begin{equation}
\overline{M} = M + KM,
\label{graymgroup}
\end{equation}
\end{itemize}
see Sec.~III.C of Ref.\ \cite{ea}.

From these conditions it follows that a stable magnetic state with the
space group $H$ can exist if 
\begin{enumerate}
\addtocounter{enumi}{2}
\item 
$H$ has at least one one-dimensional
single-valued representation $R$ following case (a) with respect to the
magnetic group $H + \{K|\frac{1}{2}\frac{1}{2}0\}H$ and case (c) with
respect to the magnetic group \mbox{$H + KH$.}
\end{enumerate}

The cases (a) and (c) are defined in Eqs.\ (7.3.45) and (7.3.47),
respectively, in the textbook of Bradley and Cracknell \cite{bc}. The
irreducible corepresentation derived from $R$ stays one-dimensional in case
(a) and becomes two-dimensional in case (c).

Within the NHM, the symmetry of the electron system is lowered as little as
possible at the transition from the paramagnetic to the magnetic phase.
Therefore  
\begin{enumerate}
\addtocounter{enumi}{3}
\item I only consider magnetic structures embedded in the symmetry of the
  tetragonal (paramagnetic) crystal.
\end{enumerate}
This means that all the Cu atoms which are connected by space group
operations in the paramagnetic phase are also connected by space group
operations in the magnetic phase.  For instance, consider the magnetic
structure (a) in Fig.~\ref{fig:strukt}.  The unit cell of this structure
contains eight Cu(2) atoms in the $A$, $A'$, $C$, and $C'$ layers and four
Cu(1) atoms in the $B$ and $B'$ layers.  The Cu(1) atoms as well as the
Cu(2) atoms are connected to each other by spatial symmetry in the
tetragonal crystal. Also in the magnetic structure, these atoms are
connected to each other by space group operations because, first, {\em
  all}~the eight Cu(2) atoms (together with their magnetic moments) in the
unit cell can be generated by application of the eight space group
operations
\begin{equation}
\begin{array}{c}
\{E|000\}, \{C_{2b}|\textstyle\frac{1}{2}\frac{1}{2}0\},
\{C_{2z}|00\frac{1}{2}\}, \{C_{2a}|\frac{1}{2}\frac{1}{2}\frac{1}{2}\},
\\\\
\{I|000\}, \{\sigma_{db}|\textstyle\frac{1}{2}\frac{1}{2}0\},
\{\sigma_z|00\frac{1}{2}\},
\{\sigma_{da}|\frac{1}{2}\frac{1}{2}\frac{1}{2}\}
\end{array}
\label{eq:4}
\end{equation}
to any Cu(2) atom.

Secondly, also all the four Cu(1) atoms in the magnetic unit cell can be
generated by applying the space group operations \gl{eq:4} to any Cu(1)
atom.  The symmetry operations \gl{eq:4} form, together with the
translations, the space group $D^{16}_{2h}$ of the magnetic structure (a).
$D^{16}_{2h}$ is a subgroup of the tetragonal space group $D^{1}_{4h}$.

Under all the space groups with the Bravais lattices $\Gamma_q$,
$\Gamma_o$, and $\Gamma_m$ there are only the four groups
\begin{equation}
  \label{eq:3}
\begin{array}{rlllll}
H_1 &=& Pnma &=& \Gamma_oD^{16}_{2h}& (62)\\
H_2 &=& Pmc2_1 &=& \Gamma_oC^{2}_{2v}& (26)\\
H_3 &=& P2_1/b &=& \Gamma_mC^{5}_{2h}& (14)\\
H_4 &=& P4/mbm &=& \Gamma_qD^5_{4h}& (127)\\
\end{array}  
\end{equation}
which satisfy the condition (iii) and belong to magnetic structures
complying with the conditions (i), (ii), and (iv). The number in
parentheses is the international number of the space group as given, e.g.,
in Table 3.7 of Ref.~\cite{bc}.

By inspection of Tables~\ref{tab:2},~\ref{tab:3},~\ref{tab:4}, and
\ref{tab:5}, respectively, it can be seen that $H_1$ (at point $S$), $H_2$
(at points $S$ and $R$), $H_3$ (at points $A$ and $E$), and $H_4$ (at
points $M$ and $A$) possess representations following condition (iii).

\section{Antiferromagnetic bands in $\text{YBa}_2\text{Cu}_3\text{O}_6$}
\label{sec:afbands}
The symmetry properties of an antiferromagnetic band in YBa$_2$Cu$_3$O$_6$
depend on both the positions of the Cu atoms and the space group of
the magnetic structure. In this section I determine all the
antiferromagnetic bands for the space groups in Eq.~\gl{eq:3} and the
related magnetic structures. The group-theoretical procedure is described,
e.g., in Ref.~\cite{la2cuo4}. The result is compared with the actual
symmetry of the Bloch functions in the band structure of
YBa$_2$Cu$_3$O$_6$.

Any antiferromagnetic band in YBa$_2$Cu$_3$O$_6$ should contain the Bloch
functions labeled by
\begin{equation}
  \label{eq:6}
\Gamma^+_5, M^+_5, Z^+_5, \text{ and } A^+_5,
\end{equation}
because these functions lie directly at the Fermi level and, hence,
particularly determine the electronic motion.


\subsection{The two magnetic structures with the space group $H_1
  = D^{16}_{2h}$ (62)} 

Both magnetic structures (a) and (b) depicted in Fig.~\ref{fig:strukt} have
the space group $H_1 = D^{16}_{2h}$ (62) with, however, different positions
of the coordinate system. Therefore, the translational parts of the space
group elements may be different. The structure (a) may be called
``fm-like'' since the coupling between $A$ and $C$ layers (and between $A'$
and $C'$ layers) is ferromagnetic.  Analogously, structure (b) may be
called ``af-like'', though the coupling between these layers is not exactly
antiferromagnetic.


\begin{figure}
\begin{center}
\includegraphics[width=.4\textwidth,angle=0]{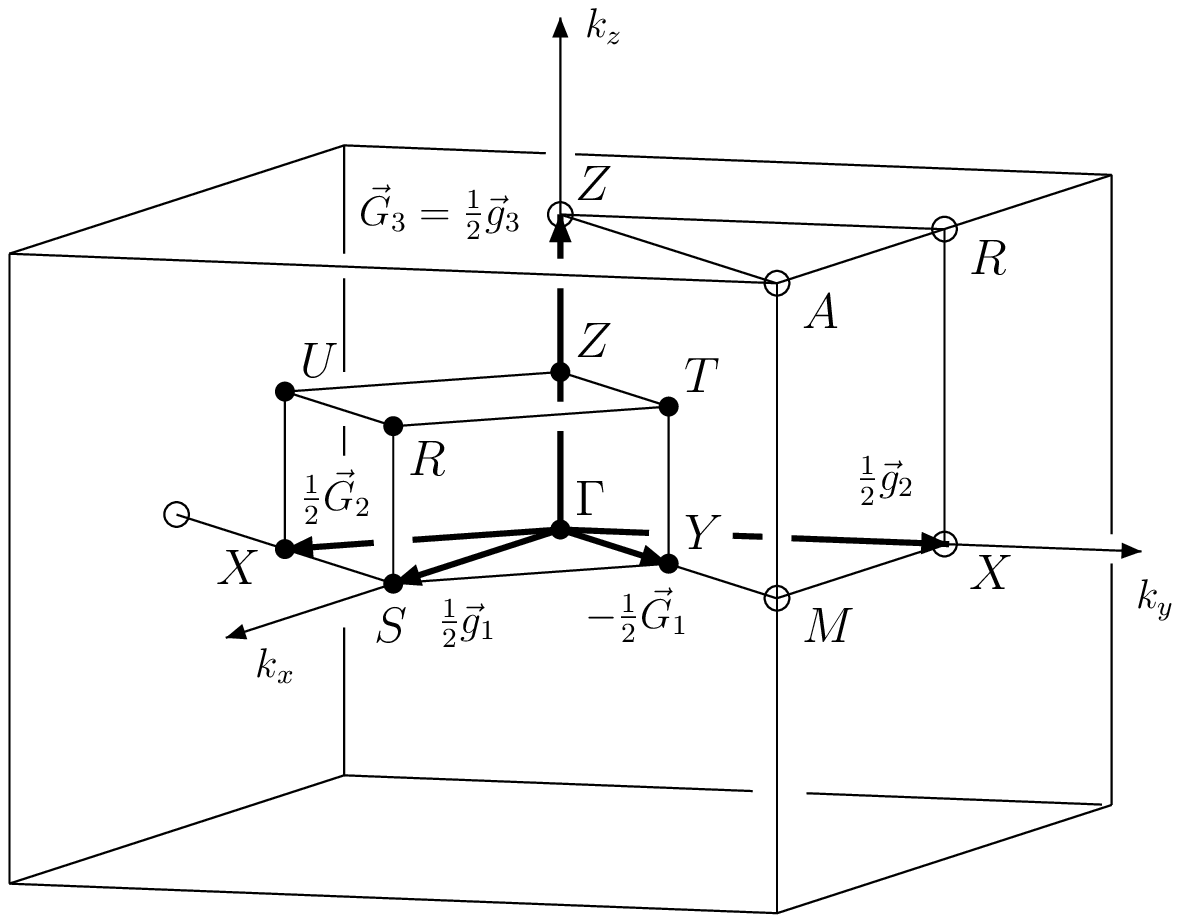}%
\begin{center}
(a)
\end{center}
\includegraphics[width=.4\textwidth,angle=0]{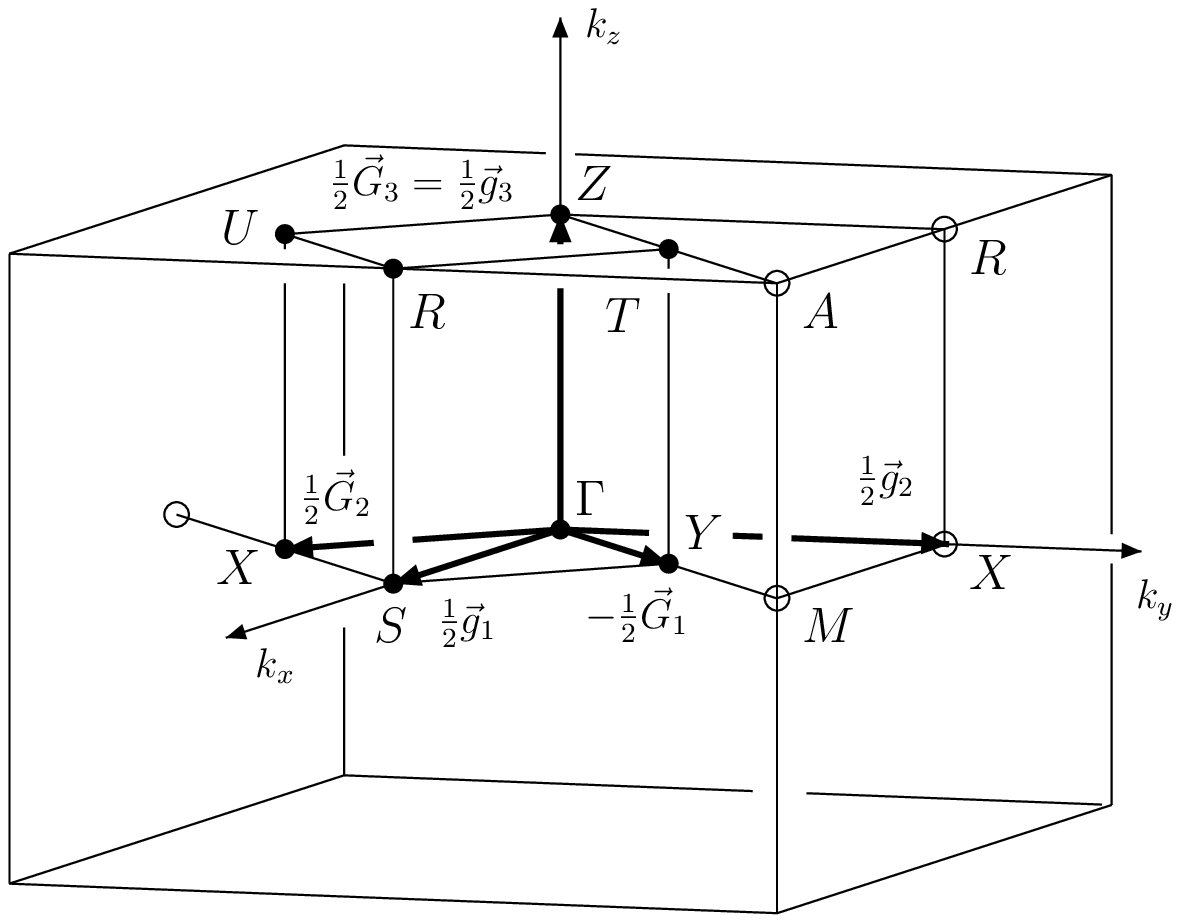}%
\begin{center}
(b)
\end{center}
\end{center}
\caption{
  Brillouin zones for paramagnetic and antiferromagnetic YBa$_2$Cu$_3$O$_6$
  with the reciprocal lattice vectors $\vec g_1, \vec g_2$, $\vec g_3$ and
  $\vec G_1, \vec G_2$, $\vec G_3$, respectively.\\
  (a) Basic domain of the Brillouin zone for the orthorhombic
  antiferromagnetic structures (a) and (b) in Fig.~\protect\ref{fig:strukt}
  within the Brillouin zone for the tetragonal paramagnetic lattice.\\
  (b) Basic domain of the Brillouin zone for the orthorhombic
  antiferromagnetic structure (d) in Fig.~\protect\ref{fig:strukt} within
  the Brillouin zone for the tetragonal paramagnetic lattice.\\
\label{fig:BZ}
}
\end{figure}


\subsubsection{The fm-like structure (a)}
Folding the conduction bands of tetragonal YBa$_2$Cu$_3$O$_6$ into the
Brillouin zone [depicted in Fig.~\ref{fig:BZ} (a)] of the fm-like
structure, the representations~\gl{eq:1} of the Bloch functions transform
as
\begin{equation}
  \label{eq:7}
\begin{array}{lclp{1cm}lcl}
\Gamma^+_5 & \rightarrow & \underline{\Gamma^+_3} +
\underline{\Gamma^+_4}&& 
\Gamma^-_5 & \rightarrow & \underline{\Gamma^-_3} +
\Gamma^-_4\\[.2cm] 
Z^+_5 & \rightarrow & \underline{\Gamma^+_1} +
\underline{\Gamma^+_2}&& 
Z^-_5 & \rightarrow & \underline{\Gamma^-_1} +
\Gamma^-_2\\[.2cm] 
M^+_3 & \rightarrow & \underline{\Gamma^+_1}&&  
M^-_4 & \rightarrow & \underline{\Gamma^-_2}\\[.2cm]  
M^+_5 & \rightarrow & \Gamma^+_3 + \underline{\Gamma^+_4}&&  
A^+_3 & \rightarrow & \underline{\Gamma^+_3}\\[.2cm]  
A^-_4 & \rightarrow & \underline{\Gamma^-_4}&&
A^+_5 & \rightarrow & \Gamma^+_1 + \underline{\Gamma^+_2}\\[.2cm]  
X^-_3 & \rightarrow & \underline{S^+_1} + \underline{S^+_2}&&  
X^+_2 & \rightarrow & \underline{S^-_3} + \underline{S^-_4}\\[.2cm]  
R^+_1 & \rightarrow & \underline{S^-_3} + \underline{S^-_4}&&  
R^+_2 & \rightarrow & \underline{S^-_1} + \underline{S^-_2}\\[.2cm]  
R^+_4 & \rightarrow & \underline{S^-_1} + \underline{S^-_2}&& 
R^-_3 & \rightarrow & \underline{S^+_3} + \underline{S^+_4},\\[.2cm]  
\end{array}
\end{equation}
see Table~\ref{tab:6}. The underlined representations form an
antiferromagnetic band, namely band 1 in Table~\ref{tab:11}. 

\subsubsection{The af-like structure (b)}
Folding the conduction bands of tetragonal YBa$_2$Cu$_3$O$_6$ into the
Brillouin zone [depicted in Fig.~\ref{fig:BZ} (a)] of the af-like
structure, the representations of the Bloch functions now transform as
\begin{equation}
  \label{eq:8}
\begin{array}{lclp{1cm}lcl}
\Gamma^+_5 & \rightarrow & \underline{\Gamma^+_3} +
\underline{\Gamma^+_4}&& 
\Gamma^-_5 & \rightarrow & \underline{\Gamma^-_3} +
\Gamma^-_4\\[.2cm] 
Z^+_5 & \rightarrow & \underline{\Gamma^-_1} +
\underline{\Gamma^-_2}&& 
Z^-_5 & \rightarrow & \Gamma^+_1 +
\underline{\Gamma^+_2}\\[.2cm] 
M^+_3 & \rightarrow & \underline{\Gamma^+_1}&&  
M^-_4 & \rightarrow & \underline{\Gamma^-_2}\\[.2cm]  
M^+_5 & \rightarrow & \Gamma^+_3 + \underline{\Gamma^+_4}&&  
A^+_3 & \rightarrow & \underline{\Gamma^-_4}\\[.2cm]  
A^-_4 & \rightarrow & \underline{\Gamma^+_3}&&
A^+_5 & \rightarrow & \underline{\Gamma^-_1} + \Gamma^-_2\\[.2cm]  
X^-_3 & \rightarrow & \underline{S^-_3} + \underline{S^-_4}&&  
X^+_2 & \rightarrow & \underline{S^+_1} + \underline{S^+_2}\\[.2cm]  
X^+_4 & \rightarrow & \underline{S^+_1} + \underline{S^+_2}&&  
X^+_1 & \rightarrow & \underline{S^+_3} + \underline{S^+_4}\\[.2cm]  
R^+_1 & \rightarrow & \underline{S^-_1} + \underline{S^-_2}&&  
R^+_2 & \rightarrow & \underline{S^-_3} + \underline{S^-_4}\\[.2cm]  
\end{array}
\end{equation}
see Table~\ref{tab:6}. The underlined representations form an
antiferromagnetic band, namely band 2 in Table~\ref{tab:12}.  

\subsection{The magnetic structure with the space group $H_2
  = C^{2}_{2v}$ (26)} 

The magnetic structure with the space group $H_2 = C^{2}_{2v}$ (26) is
depicted in Fig.~\ref{fig:strukt} (d).  Folding again the conduction bands
of tetragonal YBa$_2$Cu$_3$O$_6$ into the Brillouin zone [see 
Fig.~\ref{fig:BZ} (b)] of this structure, the representations of the Bloch
functions transform as
\begin{equation}
  \label{eq:9}
\begin{array}{lclp{1cm}lcl}
\Gamma^+_5 & \rightarrow & \underline{\Gamma_2} +
\underline{\Gamma_3}&& 
\Gamma^-_5 & \rightarrow & \Gamma_1 +
\underline{\Gamma_4}\\[.2cm] 
Z^+_5 & \rightarrow & \underline{Z_2} +
\underline{Z_3}&& 
Z^-_5 & \rightarrow & Z_1 +
\underline{Z_4}\\[.2cm] 
M^+_3 & \rightarrow & \underline{\Gamma_1}&&  
M^-_4 & \rightarrow & \underline{\Gamma_2}\\[.2cm]  
M^+_5 & \rightarrow & \Gamma_2 + \underline{\Gamma_3}&&  
A^+_3 & \rightarrow & \underline{Z_1}\\[.2cm]  
A^-_4 & \rightarrow & \underline{Z_2}&&
A^+_5 & \rightarrow & Z_2 + \underline{Z_3}\\[.2cm]  
X^-_3 & \rightarrow & \underline{S_3} + \underline{S_4}&&  
X^+_2 & \rightarrow & \underline{S_3} + \underline{S_4}\\[.2cm]  
X^+_1 & \rightarrow & \underline{S_1} + \underline{S_2}&&  
R^+_1 & \rightarrow & \underline{R_1} + \underline{R_2}\\[.2cm]  
R^+_2 & \rightarrow & \underline{R_3} + \underline{R_4}&&  
R^+_4 & \rightarrow & \underline{R_3} + \underline{R_4}\\[.2cm]  
\end{array}
\end{equation}
see Table~\ref{tab:7}. Again the underlined representations form an
antiferromagnetic band, namely band 2 in Table~\ref{tab:13}.

\subsection{The magnetic structure with the space group $H_3
  = C^{5}_{2h}$ (14)} 
The magnetic structure with the space group $H_3 = C^{5}_{2h}$ (14) is
depicted in Fig.~\ref{fig:strukt} (c).  Folding the conduction bands of
tetragonal YBa$_2$Cu$_3$O$_6$ into the Brillouin zone of this monoclinic
structure, the representations of the Bloch functions transform as
\begin{equation}
  \label{eq:10}
\begin{array}{lclp{1cm}lcl}
\Gamma^+_5 & \rightarrow & \underline{\Gamma^+_1} +
\underline{\Gamma^+_2}&& 
\Gamma^-_5 & \rightarrow & \underline{\Gamma^-_1} +
\Gamma^-_2\\[.2cm] 
Z^+_5 & \rightarrow & \underline{Z^+_1} + \underline{Z^+_2} && 
Z^-_5 & \rightarrow & \underline{Z^-_1} + Z^-_2\\[.2cm] 
M^+_3 & \rightarrow & \underline{\Gamma^+_1}&&  
M^-_4 & \rightarrow & \underline{\Gamma^-_2}\\[.2cm]  
M^+_5 & \rightarrow & \Gamma^+_1 + \underline{\Gamma^+_2}&&  
A^+_3 & \rightarrow & \underline{Z^+_1}\\[.2cm]  
A^-_4 & \rightarrow & \underline{Z^-_2}&&
A^+_5 & \rightarrow & Z^+_1 + \underline{Z^+_2}\\[.2cm]  
X^-_3 & \rightarrow & \underline{A^-_1} + \underline{A^-_2}&&  
X^+_2 & \rightarrow & \underline{A^+_1} + \underline{A^+_2}\\[.2cm]  
X^+_1 & \rightarrow & \underline{A^+_1} + \underline{A^+_2}&&  
R^+_1 & \rightarrow & \underline{E^+_1} + \underline{E^+_2}\\[.2cm]  
R^+_2 & \rightarrow & \underline{E^+_1} + \underline{E^+_2}&&  
R^-_2 & \rightarrow & \underline{E^-_1} + \underline{E^-_2}\\[.2cm]  
\end{array}
\end{equation}
see Table~\ref{tab:8}. The Brillouin zone for $\Gamma_m$ is depicted, e.g.,
in Fig.~3.3 of Ref.~\cite{bc}. Again the underlined representations form an
antiferromagnetic band, namely band 1 in Table~\ref{tab:14}.

\subsection{The magnetic structure with the space group $H_4
  = D^{5}_{4h}$ (127)} 

Putting $\alpha = \beta = 90^0$ in all the structures (a) -- (d) in
Fig.~\ref{fig:strukt}, we get a tetragonal antiferromagnetic structure with
the magnetic moments lying in $\pm z$ direction. This structure possesses
the space group $H_4 = D^{5}_{4h}$ (127). Folding the conduction bands of
tetragonal paramagnetic YBa$_2$Cu$_3$O$_6$ into the Brillouin zone of the
antiferromagnetic structure, the representations of the Bloch functions at
points $\Gamma$ and $M$ transform as
\begin{equation}
  \label{eq:11}
\begin{array}{lclp{1cm}lcl}
\Gamma^+_5 & \rightarrow & \Gamma^+_5&& 
\Gamma^-_5 & \rightarrow & \Gamma^-_5\\[.2cm] 
M^+_3 & \rightarrow & \Gamma^+_3&&  
M^-_4 & \rightarrow & \Gamma^-_4\\[.2cm]  
M^+_5 & \rightarrow & \Gamma^+_5&&  
\end{array}
\end{equation}
see Table~\ref{tab:9}. The Brillouin zone of the antiferromagnetic
structure is also the Brillouin zone for $\Gamma_q$ lying within the
Brillouin zone of the paramagnetic phase in a similar manner as depicted in
Fig.~\ref{fig:BZ} (b) for the Brillouin zone for $\Gamma_o$.
Table~\ref{tab:15} lists all the antiferromagnetic bands (in the Brillouin
zone of the antiferromagnetic structure). Not one of these bands contains
the two-dimensional representations $\Gamma^{\pm}_5$ standing on the
right-hand sites of Eq.~\gl{eq:11}. Consequently, an antiferromagnetic band
does not exist.

\subsection{Summary of Section \ref{sec:afbands}} 
All the magnetic structures (a) -- (d) in Fig.~\ref{fig:strukt} possess a
narrow partly filled antiferromagnetic band in the band structure of
YBa$_2$Cu$_3$O$_6$. The Bloch functions of these bands can be unitarily
transformed into best localized Wannier functions which are centered on the
Cu sites and symmetry-adapted to the related magnetic groups $H_i +
\{K|\textstyle\frac{1}{2}\frac{1}{2}0\}H_i$ ($i = 1, 2, 3$). Consequently,
the electrons can lower their Coulomb energy by occupying an atomiclike
state represented by these Wannier functions and, hence, all the
structures (a) -- (d) are stable in YBa$_2$Cu$_3$O$_6$.

However, the structures become unstable for $\alpha \rightarrow 90^0$ and
$\beta \rightarrow 90^0$ because for $\alpha = \beta = 90^0$ all the
structures have the space group $H_4 = D^{5}_{4h}$ (127). In this space
group, YBa$_2$Cu$_3$O$_6$ does not possess an antiferromagnetic band

Furthermore, all the structures (a) -- (d) become unstable for $\alpha
\rightarrow 0$ and $\beta \rightarrow 0$ because for $\alpha = \beta = 0$
the structures (a), (b), (c), and (d) have the space groups
$\Gamma^v_oD^{28}_{2h}$ (74), $\Gamma^v_oD^{28}_{2h}$ (74),
$\Gamma_oD^{7}_{2h}$ (53), and $\Gamma_oD^{5}_{2h}$ (51), respectively.
None of these space groups has non-real one-dimensional representations
(see, e.g., Table 5.7 of Ref.~\cite{bc}) which are necessary for
antiferromagnetic eigenstates, see point (iii) in Sec.~\ref{sec:strukt}.

Eqs.~\gl{eq:7} -- \gl{eq:10} list only the points of symmetry in the
antiferromagnetic Brillouin zone which are related to the points of
symmetry in the paramagnetic Brillouin zone. The representations at the
other points of symmetry in the antiferromagnetic Brillouin zone (which are
also given in Tables~\ref{tab:11} -- \ref{tab:14}) are related to lines and
planes of symmetry in the paramagnetic Brillouin zone. It turns out that
they are uncritical in YBa$_2$Cu$_3$O$_6$ and, hence, are not considered in
this paper for brevity.

At the points $X$ and $R$ in the paramagnetic Brillouin zone, the
antiferromagnetic bands in Eqs.~\gl{eq:7} -- \gl{eq:10} may also be chosen
in a slightly different manner. I have chosen these bands to be as narrow
as possible.

\section{Discussion}
In the framework of the NHM, all the structures (a) -- (d) in
Fig.~\ref{fig:strukt} are stable in YBa$_2$Cu$_3$O$_6$ because each of
these magnetic structures enables the electrons at the Fermi level to
occupy the energetically favorable atomiclike state. Evidently, the related
antiferromagnetic band is half-filled since YBa$_2$Cu$_3$O$_6$ is
insulating.

The structures (a) and (d) are obviously identical with the experimentally
investigated low- and high-temperature phase \cite{shamoto,tranquada},
respectively. In Figs.~1 and 3 of Refs.~\cite{tranquada} and
\cite{shamoto}, respectively, open and solid circles represent Cu sites
with mutually antiparallel magnetic moments oriented perpendicular to the
orthorhombic $c$ direction \cite{shamoto}. These open and closed circles
are also depicted in Fig.~\ref{fig:strukt} to facilitate a comparison of
the figures. It turns out that the magnetic moments belonging to open and
closed circles are not always exactly antiparallel.  Furthermore, the
magnetic moments are not exactly perpendicular to the orthorhombic $c$
direction.

To my knowledge, the af-like structure (b) has not yet been found
experimentally. Within the NHM, I have not found any reason why the fm-like
structure (a) should be more stable than the af-like structure. It is true
that the antiferromagnetic bands differ slightly at points $X$ and $R$ in
the paramagnetic Brillouin zone, see Eqs.~\gl{eq:7} and \gl{eq:8}.
However, from this difference I cannot conclude any difference between the
condensation energies of the fm-like and af-like structure. Hence, it
cannot be excluded that the af-like structure is also realized under
slightly altered conditions. It should be noted, however, that either the
fm-like or the af-like structure may exist in YBa$_2$Cu$_3$O$_6$. Any
combination of the two structures is not stable within the NHM.

Also, the structure (c) has not yet been found experimentally.  This
structure requires a monoclinic distortion of the tetragonal crystal. I
believe that such a distortion is energetically less favorable than the
orthorhombic distortions in the other structures (a), (b), and (d). Hence,
I believe that the structure (c) has a smaller condensation energy than the
other structures (a), (b), and (d).

\acknowledgements{%
  I am very indebted to Ove Jepsen for providing me with all the data I
  needed to determine the symmetry of the Bloch functions in the band
  structure of YBa$_2$Cu$_3$O$_6$. I thank Ernst Helmut Brandt for critical
  comments on the manuscript.}

\FloatBarrier
\onecolumngrid
\newpage
\appendix*
\section{Tables}
\begin{table}[h]
\caption{
  Character tables of the single-valued irreducible representations of the
  space group $P4/mmm = \Gamma_qD^{1}_{4h}$ (123) of tetragonal
  paramagnetic YBa$_2$Cu$_3$O$_6$. 
\label{tab:1}
}
\begin{tabular}[t]{lcccccccccc}
\multicolumn{11}{c}{$\Gamma (000)$, $M (\frac{1}{2}\frac{1}{2}0)$, $A
 (\frac{1}{2}\frac{1}{2}\frac{1}{2})$, and $Z (00\frac{1}{2})$}\\
 & $$ & $$ & $C^-_{4z}$ & $C_{2y}$ & $C_{2b}$ & $$ & $$ & $S^+_{4z}$ & $\sigma_y$ & $\sigma_{db}$\\
 & $E$ & $C_{2z}$ & $C^+_{4z}$ & $C_{2x}$ & $C_{2a}$ & $I$ & $\sigma_z$ & $S^-_{4z}$ & $\sigma_x$ & $\sigma_{da}$\\
\hline
$\Gamma^+_1,M^+_1,A^+_1,Z^+_1$ & 1 & 1 & 1 & 1 & 1 & 1 & 1 & 1 & 1 & 1\\
$\Gamma^+_2,M^+_2,A^+_2,Z^+_2$ & 1 & 1 & 1 & -1 & -1 & 1 & 1 & 1 & -1 & -1\\
$\Gamma^+_3,M^+_3,A^+_3,Z^+_3$ & 1 & 1 & -1 & 1 & -1 & 1 & 1 & -1 & 1 & -1\\
$\Gamma^+_4,M^+_4,A^+_4,Z^+_4$ & 1 & 1 & -1 & -1 & 1 & 1 & 1 & -1 & -1 & 1\\
$\Gamma^+_5,M^+_5,A^+_5,Z^+_5$ & 2 & -2 & 0 & 0 & 0 & 2 & -2 & 0 & 0 & 0\\
$\Gamma^-_1,M^-_1,A^-_1,Z^-_1$ & 1 & 1 & 1 & 1 & 1 & -1 & -1 & -1 & -1 & -1\\
$\Gamma^-_2,M^-_2,A^-_2,Z^-_2$ & 1 & 1 & 1 & -1 & -1 & -1 & -1 & -1 & 1 & 1\\
$\Gamma^-_3,M^-_3,A^-_3,Z^-_3$ & 1 & 1 & -1 & 1 & -1 & -1 & -1 & 1 & -1 & 1\\
$\Gamma^-_4,M^-_4,A^-_4,Z^-_4$ & 1 & 1 & -1 & -1 & 1 & -1 & -1 & 1 & 1 & -1\\
$\Gamma^-_5,M^-_5,A^-_5,Z^-_5$ & 2 & -2 & 0 & 0 & 0 & -2 & 2 & 0 & 0 & 0\\
\hline\\
\end{tabular}\hspace{5cm}
\begin{tabular}[t]{lcccccccc}
\\
\multicolumn{9}{c}{$R (0\frac{1}{2}\frac{1}{2})$ and $X
  (0\frac{1}{2}0)$}\\
& $E$ & $C_{2z}$ & $C_{2y}$ & $C_{2x}$ & $I$ & $\sigma_z$ & $\sigma_y$ & $\sigma_x$\\
\hline
$R^+_1,X^+_1$ & 1 & 1 & 1 & 1 & 1 & 1 & 1 & 1\\
$R^+_2,X^+_2$ & 1 & -1 & 1 & -1 & 1 & -1 & 1 & -1\\
$R^+_3,X^+_3$ & 1 & 1 & -1 & -1 & 1 & 1 & -1 & -1\\
$R^+_4,X^+_4$ & 1 & -1 & -1 & 1 & 1 & -1 & -1 & 1\\
$R^-_1,X^-_1$ & 1 & 1 & 1 & 1 & -1 & -1 & -1 & -1\\
$R^-_2,X^-_2$ & 1 & -1 & 1 & -1 & -1 & 1 & -1 & 1\\
$R^-_3,X^-_3$ & 1 & 1 & -1 & -1 & -1 & -1 & 1 & 1\\
$R^-_4,X^-_4$ & 1 & -1 & -1 & 1 & -1 & 1 & 1 & -1\\
\hline\\
\end{tabular}\FloatBarrier
\footnotetext{
\ \\
Note to Table~\ref{tab:1}
\begin{enumerate}
\item See the common notes to Tables~\ref{tab:1} -- \ref{tab:5} following
  Table~\ref{tab:5}.
\end{enumerate}
}
\end{table}
\begin{table}
\caption{
  Character tables and defining relations of the single-valued irreducible
  corepresentations of the magnetic group of the two orthorhombic
  antiferromagnetic structures (a) and (b) depicted in
  Fig.~\ref{fig:strukt}. 
\label{tab:2}
}
\begin{tabular}[t]{lcccccccccc}
\multicolumn{11}{c}{$\Gamma (000)$}\\
$\Gamma$, Fm: & $K$ & $\{K|\frac{1}{2}\frac{1}{2}0\}$ & $\{E|000\}$ & $\{C_{2b}|\frac{1}{2}\frac{1}{2}0\}$ & $\{C_{2z}|00\frac{1}{2}\}$ & $\{C_{2a}|\frac{1}{2}\frac{1}{2}\frac{1}{2}\}$ & $\{I|000\}$ & $\{\sigma_{db}|\frac{1}{2}\frac{1}{2}0\}$ & $\{\sigma_z|00\frac{1}{2}\}$ & $\{\sigma_{da}|\frac{1}{2}\frac{1}{2}\frac{1}{2}\}$\\
\hline
$\Gamma$, Af:& $K$ & $\{K|\frac{1}{2}\frac{1}{2}0\}$ & $\{E|000\}$ & $\{C_{2b}|\frac{1}{2}\frac{1}{2}\frac{1}{2}\}$ & $\{C_{2z}|00\frac{1}{2}\}$ & $\{C_{2a}|\frac{1}{2}\frac{1}{2}0\}$ & $\{I|00\frac{1}{2}\}$ & $\{\sigma_{db}|\frac{1}{2}\frac{1}{2}0\}$ & $\{\sigma_z|000\}$ & $\{\sigma_{da}|\frac{1}{2}\frac{1}{2}\frac{1}{2}\}$\\
\hline
$\Gamma^+_1$ & (a) & (a) & 1 & 1 & 1 & 1 & 1 & 1 & 1 & 1\\
$\Gamma^+_2$ & (a) & (a) & 1 & -1 & 1 & -1 & 1 & -1 & 1 & -1\\
$\Gamma^+_3$ & (a) & (a) & 1 & 1 & -1 & -1 & 1 & 1 & -1 & -1\\
$\Gamma^+_4$ & (a) & (a) & 1 & -1 & -1 & 1 & 1 & -1 & -1 & 1\\
$\Gamma^-_1$ & (a) & (a) & 1 & 1 & 1 & 1 & -1 & -1 & -1 & -1\\
$\Gamma^-_2$ & (a) & (a) & 1 & -1 & 1 & -1 & -1 & 1 & -1 & 1\\
$\Gamma^-_3$ & (a) & (a) & 1 & 1 & -1 & -1 & -1 & -1 & 1 & 1\\
$\Gamma^-_4$ & (a) & (a) & 1 & -1 & -1 & 1 & -1 & 1 & 1 & -1\\
\hline\\
\end{tabular}\hspace{1cm}
\begin{tabular}[t]{lcccccccccc}
\multicolumn{11}{c}{$Z (00\frac{1}{2})$, $Y (\overline{\frac{1}{2}}00)$, $X
  (0\frac{1}{2}0)$, and $U (0\frac{1}{2}\frac{1}{2})$}\\ 
$Z$, Fm: & $$ & $$ & $\{\sigma_{da}|\frac{1}{2}\frac{1}{2}\frac{1}{2}\}$ & $\{I|001\}$ & $\{C_{2a}|\frac{1}{2}\frac{1}{2}\frac{1}{2}\}$ & $$ & $$ & $\{C_{2z}|00\frac{1}{2}\}$ & $\{C_{2b}|\frac{1}{2}\frac{1}{2}1\}$ & $\{\sigma_z|00\frac{1}{2}\}$\\
$Z$, Fm: & $\{E|000\}$ & $\{E|001\}$ & $\{\sigma_{da}|\frac{1}{2}\frac{1}{2}\frac{3}{2}\}$ & $\{I|000\}$ & $\{C_{2a}|\frac{1}{2}\frac{1}{2}\frac{3}{2}\}$ & $\{\sigma_{db}|\frac{1}{2}\frac{1}{2}0\}$ & $\{\sigma_{db}|\frac{1}{2}\frac{1}{2}1\}$ & $\{C_{2z}|00\frac{3}{2}\}$ & $\{C_{2b}|\frac{1}{2}\frac{1}{2}0\}$ & $\{\sigma_z|00\frac{3}{2}\}$\\
\hline
$Z$, Af:& $$ & $$ & $\{\sigma_{da}|\frac{1}{2}\frac{1}{2}\frac{1}{2}\}$ & $\{I|00\frac{1}{2}\}$ & $\{C_{2a}|\frac{1}{2}\frac{1}{2}0\}$ & $$ & $$ & $\{C_{2z}|00\frac{1}{2}\}$ & $\{C_{2b}|\frac{1}{2}\frac{1}{2}\frac{1}{2}\}$ & $\{\sigma_z|000\}$\\
$Z$, Af: & $\{E|000\}$ & $\{E|001\}$ & $\{\sigma_{da}|\frac{1}{2}\frac{1}{2}\frac{3}{2}\}$ & $\{I|00\frac{3}{2}\}$ & $\{C_{2a}|\frac{1}{2}\frac{1}{2}1\}$ & $\{\sigma_{db}|\frac{1}{2}\frac{1}{2}0\}$ & $\{\sigma_{db}|\frac{1}{2}\frac{1}{2}1\}$ & $\{C_{2z}|00\frac{3}{2}\}$ & $\{C_{2b}|\frac{1}{2}\frac{1}{2}\frac{3}{2}\}$ & $\{\sigma_z|001\}$\\
\hline
$Y$, Fm: & $$ & $$ & $\{\sigma_{db}|\frac{1}{2}\frac{1}{2}0\}$ & $\{I|100\}$ & $\{C_{2b}|\frac{1}{2}\frac{1}{2}0\}$ & $$ & $$ & $\{C_{2a}|\frac{1}{2}\frac{1}{2}\frac{1}{2}\}$ & $\{C_{2z}|10\frac{1}{2}\}$ & $\{\sigma_{da}|\frac{1}{2}\frac{1}{2}\frac{1}{2}\}$\\
$Y$, Fm: & $\{E|000\}$ & $\{E|100\}$ & $\{\sigma_{db}|\frac{3}{2}\frac{1}{2}0\}$ & $\{I|000\}$ & $\{C_{2b}|\frac{3}{2}\frac{1}{2}0\}$ & $\{\sigma_z|00\frac{1}{2}\}$ & $\{\sigma_z|10\frac{1}{2}\}$ & $\{C_{2a}|\frac{3}{2}\frac{1}{2}\frac{1}{2}\}$ & $\{C_{2z}|00\frac{1}{2}\}$ & $\{\sigma_{da}|\frac{3}{2}\frac{1}{2}\frac{1}{2}\}$\\
\hline
$Y$, Af:& $$ & $$ & $\{\sigma_{db}|\frac{1}{2}\frac{1}{2}0\}$ & $\{I|00\frac{1}{2}\}$ & $\{C_{2b}|\frac{3}{2}\frac{1}{2}\frac{1}{2}\}$ & $$ & $$ & $\{C_{2a}|\frac{1}{2}\frac{1}{2}0\}$ & $\{C_{2z}|00\frac{1}{2}\}$ & $\{\sigma_{da}|\frac{3}{2}\frac{1}{2}\frac{1}{2}\}$\\
$Y$, Af: & $\{E|000\}$ & $\{E|100\}$ & $\{\sigma_{db}|\frac{3}{2}\frac{1}{2}0\}$ & $\{I|10\frac{1}{2}\}$ & $\{C_{2b}|\frac{1}{2}\frac{1}{2}\frac{1}{2}\}$ & $\{\sigma_z|000\}$ & $\{\sigma_z|100\}$ & $\{C_{2a}|\frac{3}{2}\frac{1}{2}0\}$ & $\{C_{2z}|10\frac{1}{2}\}$ & $\{\sigma_{da}|\frac{1}{2}\frac{1}{2}\frac{1}{2}\}$\\
\hline
$X$, Fm: & $$ & $$ & $\{\sigma_{da}|\frac{1}{2}\frac{3}{2}\frac{1}{2}\}$ & $\{I|010\}$ & $\{C_{2a}|\frac{1}{2}\frac{3}{2}\frac{1}{2}\}$ & $$ & $$ & $\{C_{2b}|\frac{1}{2}\frac{3}{2}0\}$ & $\{C_{2z}|01\frac{1}{2}\}$ & $\{\sigma_{db}|\frac{1}{2}\frac{3}{2}0\}$\\
$X$, Fm: & $\{E|000\}$ & $\{E|010\}$ & $\{\sigma_{da}|\frac{1}{2}\frac{1}{2}\frac{1}{2}\}$ & $\{I|000\}$ & $\{C_{2a}|\frac{1}{2}\frac{1}{2}\frac{1}{2}\}$ & $\{\sigma_z|00\frac{1}{2}\}$ & $\{\sigma_z|01\frac{1}{2}\}$ & $\{C_{2b}|\frac{1}{2}\frac{1}{2}0\}$ & $\{C_{2z}|00\frac{1}{2}\}$ & $\{\sigma_{db}|\frac{1}{2}\frac{1}{2}0\}$\\
\hline
$X$, Af:& $$ & $$ & $\{\sigma_{da}|\frac{1}{2}\frac{3}{2}\frac{1}{2}\}$ & $\{I|01\frac{1}{2}\}$ & $\{C_{2a}|\frac{1}{2}\frac{3}{2}0\}$ & $$ & $$ & $\{C_{2b}|\frac{1}{2}\frac{3}{2}\frac{1}{2}\}$ & $\{C_{2z}|01\frac{1}{2}\}$ & $\{\sigma_{db}|\frac{1}{2}\frac{3}{2}0\}$\\
$X$, Af: & $\{E|000\}$ & $\{E|010\}$ & $\{\sigma_{da}|\frac{1}{2}\frac{1}{2}\frac{1}{2}\}$ & $\{I|00\frac{1}{2}\}$ & $\{C_{2a}|\frac{1}{2}\frac{1}{2}0\}$ & $\{\sigma_z|000\}$ & $\{\sigma_z|010\}$ & $\{C_{2b}|\frac{1}{2}\frac{1}{2}\frac{1}{2}\}$ & $\{C_{2z}|00\frac{1}{2}\}$ & $\{\sigma_{db}|\frac{1}{2}\frac{1}{2}0\}$\\
\hline
$U$, Fm: & $$ & $$ & $\{C_{2z}|00\frac{1}{2}\}$ & $\{I|001\}$ & $\{\sigma_z|00\frac{1}{2}\}$ & $$ & $$ & $\{C_{2b}|\frac{1}{2}\frac{1}{2}0\}$ & $\{\sigma_{da}|\frac{1}{2}\frac{1}{2}\frac{3}{2}\}$ & $\{\sigma_{db}|\frac{1}{2}\frac{1}{2}0\}$\\
$U$, Fm: & $\{E|000\}$ & $\{E|001\}$ & $\{C_{2z}|00\frac{3}{2}\}$ & $\{I|000\}$ & $\{\sigma_z|00\frac{3}{2}\}$ & $\{C_{2a}|\frac{1}{2}\frac{1}{2}\frac{1}{2}\}$ & $\{C_{2a}|\frac{1}{2}\frac{1}{2}\frac{3}{2}\}$ & $\{C_{2b}|\frac{1}{2}\frac{1}{2}1\}$ & $\{\sigma_{da}|\frac{1}{2}\frac{1}{2}\frac{1}{2}\}$ & $\{\sigma_{db}|\frac{1}{2}\frac{1}{2}1\}$\\
\hline
$U$, Af:& $$ & $$ & $\{C_{2z}|00\frac{1}{2}\}$ & $\{I|00\frac{1}{2}\}$ & $\{\sigma_z|000\}$ & $$ & $$ & $\{C_{2b}|\frac{1}{2}\frac{1}{2}\frac{3}{2}\}$ & $\{\sigma_{da}|\frac{1}{2}\frac{1}{2}\frac{3}{2}\}$ & $\{\sigma_{db}|\frac{1}{2}\frac{1}{2}0\}$\\
$U$, Af: & $\{E|000\}$ & $\{E|001\}$ & $\{C_{2z}|00\frac{3}{2}\}$ & $\{I|00\frac{3}{2}\}$ & $\{\sigma_z|001\}$ & $\{C_{2a}|\frac{1}{2}\frac{1}{2}0\}$ & $\{C_{2a}|\frac{1}{2}\frac{1}{2}1\}$ & $\{C_{2b}|\frac{1}{2}\frac{1}{2}\frac{1}{2}\}$ & $\{\sigma_{da}|\frac{1}{2}\frac{1}{2}\frac{1}{2}\}$ & $\{\sigma_{db}|\frac{1}{2}\frac{1}{2}1\}$\\
\hline
$Z_1, Y_1, X_1, U_1$ & 2 & -2 & 0 & 0 & 0 & 2 & -2 & 0 & 0 & 0\\
$Z_2, Y_2, X_2, U_2$ & 2 & -2 & 0 & 0 & 0 & -2 & 2 & 0 & 0 & 0\\
\hline\\
\end{tabular}\hspace{1cm}
\begin{tabular}[t]{lcccccccccc}
\multicolumn{11}{c}{$R
  (\overline{\frac{1}{2}}\frac{1}{2}\overline{\frac{1}{2}})$ and $T
  (\overline{\frac{1}{2}}0\overline{\frac{1}{2}})$}\\ 
$R$, Fm: & $$ & $$ & $$ & $$ & $\{\sigma_z|00\frac{3}{2}\}$ & $\{C_{2a}|\frac{1}{2}\frac{1}{2}\frac{1}{2}\}$ & $\{I|001\}$ & $\{C_{2b}|\frac{1}{2}\frac{1}{2}0\}$ & $\{C_{2z}|00\frac{3}{2}\}$ & $\{\sigma_{da}|\frac{1}{2}\frac{1}{2}\frac{1}{2}\}$\\
$R$, Fm: & $\{E|000\}$ & $\{\sigma_{db}|\frac{1}{2}\frac{1}{2}1\}$ & $\{E|001\}$ & $\{\sigma_{db}|\frac{1}{2}\frac{1}{2}0\}$ & $\{\sigma_z|00\frac{1}{2}\}$ & $\{C_{2a}|\frac{1}{2}\frac{1}{2}\frac{3}{2}\}$ & $\{I|000\}$ & $\{C_{2b}|\frac{1}{2}\frac{1}{2}1\}$ & $\{C_{2z}|00\frac{1}{2}\}$ & $\{\sigma_{da}|\frac{1}{2}\frac{1}{2}\frac{3}{2}\}$\\
\hline
$R$, Af:& $$ & $$ & $$ & $$ & $\{\sigma_z|001\}$ & $\{C_{2a}|\frac{1}{2}\frac{1}{2}0\}$ & $\{I|00\frac{3}{2}\}$ & $\{C_{2b}|\frac{1}{2}\frac{1}{2}\frac{1}{2}\}$ & $\{C_{2z}|00\frac{1}{2}\}$ & $\{\sigma_{da}|\frac{1}{2}\frac{1}{2}\frac{3}{2}\}$\\
$R$, Af: & $\{E|000\}$ & $\{\sigma_{db}|\frac{1}{2}\frac{1}{2}1\}$ & $\{E|001\}$ & $\{\sigma_{db}|\frac{1}{2}\frac{1}{2}0\}$ & $\{\sigma_z|000\}$ & $\{C_{2a}|\frac{1}{2}\frac{1}{2}1\}$ & $\{I|00\frac{1}{2}\}$ & $\{C_{2b}|\frac{1}{2}\frac{1}{2}\frac{3}{2}\}$ & $\{C_{2z}|00\frac{3}{2}\}$ & $\{\sigma_{da}|\frac{1}{2}\frac{1}{2}\frac{1}{2}\}$\\
\hline
$T$, Fm: & $$ & $$ & $$ & $$ & $\{\sigma_z|00\frac{3}{2}\}$ & $\{\sigma_{db}|\frac{1}{2}\frac{1}{2}0\}$ & $\{I|001\}$ & $\{\sigma_{da}|\frac{1}{2}\frac{1}{2}\frac{1}{2}\}$ & $\{C_{2z}|00\frac{3}{2}\}$ & $\{C_{2b}|\frac{1}{2}\frac{1}{2}0\}$\\
$T$, Fm: & $\{E|000\}$ & $\{C_{2a}|\frac{1}{2}\frac{1}{2}\frac{3}{2}\}$ & $\{E|001\}$ & $\{C_{2a}|\frac{1}{2}\frac{1}{2}\frac{1}{2}\}$ & $\{\sigma_z|00\frac{1}{2}\}$ & $\{\sigma_{db}|\frac{1}{2}\frac{1}{2}1\}$ & $\{I|000\}$ & $\{\sigma_{da}|\frac{1}{2}\frac{1}{2}\frac{3}{2}\}$ & $\{C_{2z}|00\frac{1}{2}\}$ & $\{C_{2b}|\frac{1}{2}\frac{1}{2}1\}$\\
\hline
$T$, Af:& $$ & $$ & $$ & $$ & $\{\sigma_z|001\}$ & $\{\sigma_{db}|\frac{1}{2}\frac{1}{2}0\}$ & $\{I|00\frac{3}{2}\}$ & $\{\sigma_{da}|\frac{1}{2}\frac{1}{2}\frac{3}{2}\}$ & $\{C_{2z}|00\frac{1}{2}\}$ & $\{C_{2b}|\frac{1}{2}\frac{1}{2}\frac{1}{2}\}$\\
$T$, Af: & $\{E|000\}$ & $\{C_{2a}|\frac{1}{2}\frac{1}{2}1\}$ & $\{E|001\}$ & $\{C_{2a}|\frac{1}{2}\frac{1}{2}0\}$ & $\{\sigma_z|000\}$ & $\{\sigma_{db}|\frac{1}{2}\frac{1}{2}1\}$ & $\{I|00\frac{1}{2}\}$ & $\{\sigma_{da}|\frac{1}{2}\frac{1}{2}\frac{1}{2}\}$ & $\{C_{2z}|00\frac{3}{2}\}$ & $\{C_{2b}|\frac{1}{2}\frac{1}{2}\frac{3}{2}\}$\\
\hline
$R_1, T_1$ & 2 & 2i & -2 & -2i & 0 & 0 & 0 & 0 & 0 & 0\\
$R_2, T_2$ & 2 & -2i & -2 & 2i & 0 & 0 & 0 & 0 & 0 & 0\\
\hline\\
\end{tabular}\hspace{1cm}
\begin{center}
\begin{tabular}[t]{lcc}
\multicolumn{3}{c}{$Z$, $Y$, and $X$}\\  
& $K$ & $\{K|\frac{1}{2}\frac{1}{2}0\}$\\
\hline
$Z_1$, $Z_2$, $Y_1$, $Y_2$, $X_1$, $X_2$ & (a) & (a) \\  
\hline
\end{tabular}\hspace{.5cm}
\begin{tabular}[t]{lcc}
\multicolumn{3}{c}{$U$}\\ 
& $K$ & $\{K|\frac{1}{2}\frac{1}{2}0\}$\\
\hline
$U_1$, $U_2$ & (a) & (c) \\  
\hline
\end{tabular}\hspace{.5cm}
\begin{tabular}[t]{lcc}
\multicolumn{3}{c}{$R$}\\  
& $K$ & $\{K|\frac{1}{2}\frac{1}{2}0\}$\\
\hline
$R_1$, $R_2$ & (c) & (a) \\  
\hline
\end{tabular}\hspace{.5cm}
\begin{tabular}[t]{lcc}
\multicolumn{3}{c}{$T$}\\ 
& $K$ & $\{K|\frac{1}{2}\frac{1}{2}0\}$\\
\hline
$T_1$, $T_2$ & (c) & (c) \\  
\hline\\
\end{tabular}
\end{center}
\end{table}
\begin{table}
\addtocounter{table}{-1}
\caption{continued}
\begin{tabular}[t]{lccccccccc}
\multicolumn{10}{c}{$S (\overline{\frac{1}{2}}\frac{1}{2}0)$}\\
$S$, Fm: & $K$ & $\{K|\frac{1}{2}\frac{1}{2}0\}$ & $\{E|000\}$ & $\{C_{2a}|\frac{1}{2}\frac{3}{2}\frac{1}{2}\}$ & $\{E|010\}$ & $\{C_{2a}|\frac{1}{2}\frac{1}{2}\frac{1}{2}\}$ & $\{C_{2z}|00\frac{1}{2}\}$ & $\{C_{2b}|\frac{1}{2}\frac{3}{2}0\}$ & $\{C_{2z}|01\frac{1}{2}\}$\\
\hline
$S$, Af:& $K$ & $\{K|\frac{1}{2}\frac{1}{2}0\}$ & $\{E|000\}$ & $\{C_{2a}|\frac{1}{2}\frac{3}{2}0\}$ & $\{E|010\}$ & $\{C_{2a}|\frac{1}{2}\frac{1}{2}0\}$ & $\{C_{2z}|01\frac{1}{2}\}$ & $\{C_{2b}|\frac{1}{2}\frac{1}{2}\frac{1}{2}\}$ & $\{C_{2z}|00\frac{1}{2}\}$\\
\hline
$S^-_1$ & (c) & (a) & 1 & i & -1 & -i & 1 & i & -1\\
$S^-_2$ & (c) & (a) & 1 & -i & -1 & i & 1 & -i & -1\\
$S^-_3$ & (c) & (a) & 1 & i & -1 & -i & -1 & -i & 1\\
$S^-_4$ & (c) & (a) & 1 & -i & -1 & i & -1 & i & 1\\
$S^+_1$ & (c) & (a) & 1 & i & -1 & -i & 1 & i & -1\\
$S^+_2$ & (c) & (a) & 1 & -i & -1 & i & 1 & -i & -1\\
$S^+_3$ & (c) & (a) & 1 & i & -1 & -i & -1 & -i & 1\\
$S^+_4$ & (c) & (a) & 1 & -i & -1 & i & -1 & i & 1\\
\hline\\
\end{tabular}\hspace{1cm}
\begin{tabular}[t]{lccccccccc}
\multicolumn{10}{c}{$S (\overline{\frac{1}{2}}\frac{1}{2}0)$\qquad $(continued)$}\\
$S$, Fm: & $\{C_{2b}|\frac{1}{2}\frac{1}{2}0\}$ & $\{I|000\}$ & $\{\sigma_{da}|\frac{1}{2}\frac{3}{2}\frac{1}{2}\}$ & $\{I|010\}$ & $\{\sigma_{da}|\frac{1}{2}\frac{1}{2}\frac{1}{2}\}$ & $\{\sigma_z|00\frac{1}{2}\}$ & $\{\sigma_{db}|\frac{1}{2}\frac{3}{2}0\}$ & $\{\sigma_z|01\frac{1}{2}\}$ & $\{\sigma_{db}|\frac{1}{2}\frac{1}{2}0\}$\\
\hline
$S$, Af:& $\{C_{2b}|\frac{1}{2}\frac{3}{2}\frac{1}{2}\}$ & $\{I|01\frac{1}{2}\}$ & $\{\sigma_{da}|\frac{1}{2}\frac{1}{2}\frac{1}{2}\}$ & $\{I|00\frac{1}{2}\}$ & $\{\sigma_{da}|\frac{1}{2}\frac{3}{2}\frac{1}{2}\}$ & $\{\sigma_z|000\}$ & $\{\sigma_{db}|\frac{1}{2}\frac{3}{2}0\}$ & $\{\sigma_z|010\}$ & $\{\sigma_{db}|\frac{1}{2}\frac{1}{2}0\}$\\
\hline
$S^-_1$ & -i & 1 & i & -1 & -i & 1 & i & -1 & -i\\
$S^-_2$ & i & 1 & -i & -1 & i & 1 & -i & -1 & i\\
$S^-_3$ & i & 1 & i & -1 & -i & -1 & -i & 1 & i\\
$S^-_4$ & -i & 1 & -i & -1 & i & -1 & i & 1 & -i\\
$S^+_1$ & -i & -1 & -i & 1 & i & -1 & -i & 1 & i\\
$S^+_2$ & i & -1 & i & 1 & -i & -1 & i & 1 & -i\\
$S^+_3$ & i & -1 & -i & 1 & i & 1 & i & -1 & -i\\
$S^+_4$ & -i & -1 & i & 1 & -i & 1 & -i & -1 & i\\
\hline\\
\end{tabular}\hspace{1cm}
\footnotetext{
\ \\
Notes to Table~\ref{tab:2}
\begin{enumerate}
\item Both the fm-like antiferromagnetic structure (a) and the af-like
  antiferromagnetic structure (b) in Fig.~\ref{fig:strukt} possess the
  space group $H = Pnma = \Gamma_oD^{16}_{2h}$ (62) with, however,
  different positions of the origin. Hence, the space group elements may
  differ in the nonprimitive translations.
\item $K$ stands for the operator of time-inversion. $M = D^{16}_{2h} +
  \{K|\frac{1}{2}\frac{1}{2}0\}D^{16}_{2h}$ is the magnetic group of the
  antiferromagnetic structure.
\item Following the label Fm or Af, the upper rows list the space group
  elements of the fm-like and af-like structure, respectively, of the
  little group of the point of symmetry given at the beginning of the row.
  In the tables for $\Gamma$ and $S$, the two anti-unitary elements $K$ and
  $\{K|\frac{1}{2}\frac{1}{2}0\}$ are also included in these upper
  rows.
\item The space group elements of $D^{16}_{2h}$ used in this paper differ
  from those given by Bradley and Cracknell \protect\cite{bc} because the
  orientation of both the coordinate axis and the basic translations is
  different from that used by Bradley and Cracknell. As a consequence, our
  vectors of the reciprocal lattice have a different orientation, too. This
  is irrelevant within this paper since all the tables are consistent with
  our coordinate systems. However, the entries for the space group
  $D^{16}_{2h}$ in Table 5.7 of Bradley and Cracknell \protect\cite{bc}
  cannot be compared with this Table~\ref{tab:2} before the points of
  symmetry are renamed as follows:
\begin{center} 
\begin{tabular}{lccccccccc}
Name used in Table 5.7 of
Ref.~\protect\cite{bc}&\ \ & 
$\Gamma$&$Y$&$X$&$Z$&$U$&$T$&$S$&$R$\\
\hline
Name used in this Table~\ref{tab:2}&&$\Gamma$&$Z$&$Y$&$X$&$S$&$U$&$T$&$R$\\
\end{tabular}
\end{center}
\item The antiferromagnetic state
  belongs to a corepresentation of $M$ which is derived from one of the
  representations at point $S$ because the little group of $S$ comprises
  the whole space group and has one-dimensional representations following
  case $(c)$ and case $(a)$ with respect to $K$ and
  $\{K|\frac{1}{2}\frac{1}{2}0\}$, respectively.
\item See also the common notes to Tables~\ref{tab:1} -- \ref{tab:5}
  following Table~\ref{tab:5}. 
\end{enumerate}
}
\end{table}
\begin{table}
\caption{
  Character tables and defining relations of the single-valued irreducible
  corepresentations of the magnetic group of the orthorhombic
  antiferromagnetic structure (d) depicted in Fig.~\ref{fig:strukt}.
\label{tab:3}
}
\begin{tabular}[t]{lcccc}
\multicolumn{5}{c}{$\Gamma (000)$, $Z (00\frac{1}{2})$, $X (0\frac{1}{2}0)$
  and $U (0\overline{\frac{1}{2}}\frac{1}{2})$}\\ 
$\Gamma, Z:$ & $\{E|000\}$ & $\{C_{2a}|\frac{1}{2}\frac{1}{2}0\}$ & $\{\sigma_{db}|\frac{1}{2}\frac{1}{2}0\}$ & $\{\sigma_z|000\}$\\
\hline
$X$: & $\{E|000\}$ & $\{C_{2a}|\frac{1}{2}\frac{3}{2}0\}$ & $\{\sigma_{db}|\frac{1}{2}\frac{3}{2}0\}$ & $\{\sigma_z|000\}$\\
\hline
$U$: & $\{E|000\}$ & $\{C_{2a}|\frac{1}{2}\frac{1}{2}1\}$ & $\{\sigma_{db}|\frac{1}{2}\frac{1}{2}1\}$ & $\{\sigma_z|000\}$\\
\hline
$\Gamma_1, Z_1, X_1, U_1$ & 1 & 1 & 1 & 1\\
$\Gamma_3, Z_3, X_3, U_3$ & 1 & 1 & -1 & -1\\
$\Gamma_2, Z_2, X_2, U_2$ & 1 & -1 & 1 & -1\\
$\Gamma_4, Z_4, X_4, U_4$ & 1 & -1 & -1 & 1\\
\hline\\
\end{tabular}\hspace{1cm}
\begin{tabular}[t]{lcccccccc}
\multicolumn{9}{c}{$Y (\overline{\frac{1}{2}}00)$, $T
 (\overline{\frac{1}{2}}0\frac{1}{2})$, $S
 (\overline{\frac{1}{2}}\frac{1}{2}0)$ and $R 
 (\overline{\frac{1}{2}}\overline{\frac{1}{2}}\frac{1}{2})$}\\ 
$Y$: & $\{E|000\}$ & $\{C_{2a}|\frac{3}{2}\frac{1}{2}0\}$ & $\{E|100\}$ & $\{C_{2a}|\frac{1}{2}\frac{1}{2}0\}$ & $\{\sigma_z|000\}$ & $\{\sigma_{db}|\frac{3}{2}\frac{1}{2}0\}$ & $\{\sigma_z|100\}$ & $\{\sigma_{db}|\frac{1}{2}\frac{1}{2}0\}$\\
\hline
$T$: & $\{E|000\}$ & $\{C_{2a}|\frac{1}{2}\frac{1}{2}1\}$ & $\{E|001\}$ & $\{C_{2a}|\frac{1}{2}\frac{1}{2}0\}$ & $\{\sigma_z|000\}$ & $\{\sigma_{db}|\frac{1}{2}\frac{1}{2}1\}$ & $\{\sigma_z|001\}$ & $\{\sigma_{db}|\frac{1}{2}\frac{1}{2}0\}$\\
\hline
$S$: & $\{E|000\}$ & $\{C_{2a}|\frac{1}{2}\frac{1}{2}0\}$ & $\{E|010\}$ & $\{C_{2a}|\frac{1}{2}\frac{3}{2}0\}$ & $\{\sigma_z|000\}$ & $\{\sigma_{db}|\frac{1}{2}\frac{1}{2}0\}$ & $\{\sigma_z|010\}$ & $\{\sigma_{db}|\frac{1}{2}\frac{3}{2}0\}$\\
\hline
$R$: & $\{E|000\}$ & $\{C_{2a}|\frac{1}{2}\frac{1}{2}0\}$ & $\{E|001\}$ & $\{C_{2a}|\frac{1}{2}\frac{1}{2}1\}$ & $\{\sigma_z|000\}$ & $\{\sigma_{db}|\frac{1}{2}\frac{1}{2}0\}$ & $\{\sigma_z|001\}$ & $\{\sigma_{db}|\frac{1}{2}\frac{1}{2}1\}$\\
\hline
$Y_1, T_1, S_1, R_1$ & 1 & i & -1 & -i & 1 & i & -1 & -i\\
$Y_2, T_2, S_2, R_2$ & 1 & -i & -1 & i & 1 & -i & -1 & i\\
$Y_3, T_3, S_3, R_3$ & 1 & i & -1 & -i & -1 & -i & 1 & i\\
$Y_4, T_4, S_4, R_4$ & 1 & -i & -1 & i & -1 & i & 1 & -i\\
\hline\\
\end{tabular}\hspace{1cm}
\begin{tabular}[t]{lcc}
\multicolumn{3}{c}{$\Gamma$ and $Z$}\\ 
& $K$ & $\{K|\frac{1}{2}\frac{1}{2}0\}$\\
\hline
$\Gamma_1$ - $\Gamma_4, Z_1$ - $Z_4$ & (a) & (a)\\
\hline\\
\end{tabular}\hspace{1cm}
\begin{tabular}[t]{lcc}
\multicolumn{3}{c}{$X$ and $U$}\\ 
& $K$ & $\{K|\frac{1}{2}\frac{1}{2}0\}$\\
\hline
$X_1$ - $X_4, U_1$ - $U_4$ & (a) & (c)\\
\hline\\
\end{tabular}\hspace{1cm}
\begin{tabular}[t]{lcc}
\multicolumn{3}{c}{$Y$ and $T$}\\ 
& $K$ & $\{K|\frac{1}{2}\frac{1}{2}0\}$\\
\hline
$Y_1$ - $Y_4, T_1$ - $T_4$ & (c) & (c)\\
\hline\\
\end{tabular}\hspace{1cm}
\begin{tabular}[t]{lcc}
\multicolumn{3}{c}{$S$ and $R$}\\ 
& $K$ & $\{K|\frac{1}{2}\frac{1}{2}0\}$\\
\hline
$S_1$ - $S_4, R_1$ - $R_4$ & (c) & (a)\\
\hline\\
\end{tabular}\hspace{1cm}
\footnotetext{
\ \\
Notes to Table~\ref{tab:3}
\begin{enumerate}
\item The antiferromagnetic structure (d) in Fig.~\ref{fig:strukt}
  possesses the space group $H = Pmc2_1 = \Gamma_oC^{2}_{2v}$ (26).
\item $K$ stands for the operator of time-inversion. $M = C^{2}_{2v} +
  \{K|\frac{1}{2}\frac{1}{2}0\}C^{2}_{2v}$ is the magnetic group of the
  antiferromagnetic structure.
\item The upper rows list the space group elements of the little group of
  the point(s) of symmetry given at the beginning of the row.
\item The space group elements of $C^{2}_{2v}$ used in this paper differ
  from those given by Bradley and Cracknell \protect\cite{bc} because the
  orientation of both the coordinate axis and the basic translations is
  different from that used by Bradley and Cracknell. As a consequence, our
  vectors of the reciprocal lattice have a different orientation, too. This
  is irrelevant within this paper since all the tables are consistent with
  our coordinate systems. However, the entries for the space group
  $C^{2}_{2v}$ in Table 5.7 of Bradley and Cracknell \protect\cite{bc}
  cannot be compared with this Table~\ref{tab:3} before the points of
  symmetry are renamed as follows:
\begin{center} 
\begin{tabular}{lccccccccc}
Name used in Table 5.7 of Ref.~\protect\cite{bc}&\ \ &
$\Gamma$&$Y$&$X$&$Z$&$U$&$T$&$S$&$R$\\
\hline
Name used in this Table~\ref{tab:3}&&
$\Gamma$&$X$&$Z$&$Y$&$T$&$S$&$U$&$R$\\
\end{tabular}
\end{center}
\item The antiferromagnetic state belongs to a corepresentation of $M$
  which is derived from one of the representations at points $S$ or $R$
  because the little groups of $S$ and $R$ comprise the whole space group
  and have one-dimensional representations following case $(c)$ and case
  $(a)$ with respect to $K$ and $\{K|\frac{1}{2}\frac{1}{2}0\}$,
  respectively.
\item See also the common notes to Tables~\ref{tab:1} -- \ref{tab:5}
  following Table~\ref{tab:5}. 
\end{enumerate}
}
\end{table}
\begin{table}
\caption{
  Character tables and defining relations of the single-valued irreducible
  corepresentations of the magnetic group of the monoclinic
  antiferromagnetic structure depicted (c) in Fig.~\ref{fig:strukt}.  
\label{tab:4}
}
\begin{tabular}[t]{lcccccc}
\multicolumn{7}{c}{$\Gamma (000)$ and $Z (00\overline{\frac{1}{2}})$}\\
$\Gamma, Z:$ & $K$ & $\{K|\frac{1}{2}\frac{1}{2}0\}$ & $\{E|000\}$ &
 $\{C_{2b}|\frac{1}{2}\frac{1}{2}0\}$ & $\{I|000\}$ &
 $\{\sigma_{db}|\frac{1}{2}\frac{1}{2}0\}$\\ 
\hline
$\Gamma^+_1, Z^+_1$ & (a) & (a) & 1 & 1 & 1 & 1\\
$\Gamma^-_1, Z^-_1$ & (a) & (a) & 1 & 1 & -1 & -1\\
$\Gamma^+_2, Z^+_2$ & (a) & (a) & 1 & -1 & 1 & -1\\
$\Gamma^-_2, Z^-_2$ & (a) & (a) & 1 & -1 & -1 & 1\\
\hline\\
\end{tabular}\hspace{3cm}
\begin{tabular}[t]{lccccccc}
\multicolumn{8}{c}{$B (\overline{\frac{1}{2}}00)$, $Y (0\frac{1}{2}0)$, $C
  (0\frac{1}{2}\overline{\frac{1}{2}})$, and $D
  (\frac{1}{2}0\overline{\frac{1}{2}})$}\\ 
$B:$ &  &  & $$ & $$ & $\{\sigma_{db}|\frac{3}{2}\frac{1}{2}0\}$ & $\{I|000\}$ & $\{C_{2b}|\frac{1}{2}\frac{1}{2}0\}$\\
$B:$ & $K$ & $\{K|\frac{1}{2}\frac{1}{2}0\}$ & $\{E|000\}$ & $\{E|100\}$ & $\{\sigma_{db}|\frac{1}{2}\frac{1}{2}0\}$ & $\{I|100\}$ & $\{C_{2b}|\frac{3}{2}\frac{1}{2}0\}$\\
\hline
$Y:$ &  &  & $$ & $$ & $\{C_{2b}|\frac{1}{2}\frac{3}{2}0\}$ & $\{I|000\}$ & $\{\sigma_{db}|\frac{1}{2}\frac{1}{2}0\}$\\
$Y:$ & $K$ & $\{K|\frac{1}{2}\frac{1}{2}0\}$ & $\{E|000\}$ & $\{E|010\}$ & $\{C_{2b}|\frac{1}{2}\frac{1}{2}0\}$ & $\{I|010\}$ & $\{\sigma_{db}|\frac{1}{2}\frac{3}{2}0\}$\\
\hline
$C:$ &  &  & $$ & $$ & $\{C_{2b}|\frac{1}{2}\frac{1}{2}1\}$ & $\{I|000\}$ & $\{\sigma_{db}|\frac{1}{2}\frac{1}{2}0\}$\\
$C:$ & $K$ & $\{K|\frac{1}{2}\frac{1}{2}0\}$ & $\{E|000\}$ & $\{E|001\}$ & $\{C_{2b}|\frac{1}{2}\frac{1}{2}0\}$ & $\{I|001\}$ & $\{\sigma_{db}|\frac{1}{2}\frac{1}{2}1\}$\\
\hline
$D:$ &  &  & $$ & $$ & $\{\sigma_{db}|\frac{1}{2}\frac{1}{2}1\}$ & $\{I|000\}$ & $\{C_{2b}|\frac{1}{2}\frac{1}{2}0\}$\\
$D:$ & $K$ & $\{K|\frac{1}{2}\frac{1}{2}0\}$ & $\{E|000\}$ & $\{E|001\}$ & $\{\sigma_{db}|\frac{1}{2}\frac{1}{2}0\}$ & $\{I|001\}$ & $\{C_{2b}|\frac{1}{2}\frac{1}{2}1\}$\\
\hline
$B_1, Y_1, C_1, D_1$ & (a) & (a) & 2 & -2 & 0 & 0 & 0\\
\hline\\
\end{tabular}\hspace{1cm}
\begin{tabular}[t]{lcccccccccc}
\multicolumn{11}{c}{$A (\overline{\frac{1}{2}}\frac{1}{2}0)$ and $E
  (\frac{1}{2}\frac{1}{2}\overline{\frac{1}{2}})$}\\ 
$A:$ & $K$ & $\{K|\frac{1}{2}\frac{1}{2}0\}$ & $\{E|000\}$ & $\{C_{2b}|\frac{1}{2}\frac{1}{2}0\}$ & $\{E|010\}$ & $\{C_{2b}|\frac{1}{2}\frac{3}{2}0\}$ & $\{I|000\}$ & $\{\sigma_{db}|\frac{1}{2}\frac{1}{2}0\}$ & $\{I|010\}$ & $\{\sigma_{db}|\frac{1}{2}\frac{3}{2}0\}$\\
\hline
$E:$ & $K$ & $\{K|\frac{1}{2}\frac{1}{2}0\}$ & $\{E|000\}$ & $\{C_{2b}|\frac{1}{2}\frac{1}{2}0\}$ & $\{E|001\}$ & $\{C_{2b}|\frac{1}{2}\frac{1}{2}1\}$ & $\{I|000\}$ & $\{\sigma_{db}|\frac{1}{2}\frac{1}{2}0\}$ & $\{I|001\}$ & $\{\sigma_{db}|\frac{1}{2}\frac{1}{2}1\}$\\
\hline
$A^+_1, E^+_1$ & (c) & (a) & 1 & i & -1 & -i & 1 & i & -1 & -i\\
$A^+_2, E^+_2$ & (c) & (a) & 1 & -i & -1 & i & 1 & -i & -1 & i\\
$A^-_1, E^-_1$ & (c) & (a) & 1 & i & -1 & -i & -1 & -i & 1 & i\\
$A^-_2, E^-_2$ & (c) & (a) & 1 & -i & -1 & i & -1 & i & 1 & -i\\
\hline\\
\end{tabular}\hspace{1cm}
\footnotetext{
\ \\
Notes to Table~\ref{tab:4}
\begin{enumerate}
\item The antiferromagnetic structure (c) in Fig.~\ref{fig:strukt}
  possesses the space group $H = P2_1/b = \Gamma_mC^{5}_{2h}$ (14).
\item $K$ stands for the operator of time-inversion. $M = C^{5}_{2h} +
  \{K|\frac{1}{2}\frac{1}{2}0\}C^{5}_{2h}$ is the magnetic group of the
  antiferromagnetic structure.
\item The upper rows list the two anti-unitary elements $K$ and
  $\{K|\frac{1}{2}\frac{1}{2}0\}$ and the space group elements of the
  little group of the point(s) of symmetry given at the beginning of the
  row.
\item The space group elements of $C^{5}_{2h}$ used in this paper differ
  from those given by Bradley and Cracknell \protect\cite{bc} because the
  orientation of both the coordinate axis and the basic translations is
  different from that used by Bradley and Cracknell. As a consequence, our
  vectors of the reciprocal lattice have a different orientation, too. This
  is irrelevant within this paper since all the tables are consistent with
  our coordinate systems. However, the entries for the space group
  $C^{5}_{2h}$ in Table 5.7 of Bradley and Cracknell \protect\cite{bc}
  cannot be compared with this Table~\ref{tab:4} before the points of
  symmetry are renamed as follows:
\begin{center} 
\begin{tabular}{lccccccccc}
Name used in Table 5.7 of Ref.~\protect\cite{bc}&\ \ &
$\Gamma$&$B$&$Y$&$Z$&$C$&$D$&$A$&$E$\\
\hline
Name used in this Table~\ref{tab:4}&&
$\Gamma$&$B$&$Z$&$Y$&$C$&$A$&$D$&$E$\\
\end{tabular}
\end{center}
\item The antiferromagnetic state, if
  it exists, belongs to a corepresentation of $M$ which is derived from
  one of the representations at points $A$ or $E$ because the little groups
  of $A$ and $E$ comprise the whole space group and have one-dimensional
  representations following case $(c)$ and case $(a)$ with respect to $K$
  and $\{K|\frac{1}{2}\frac{1}{2}0\}$, respectively.
\item See also the common notes to Tables~\ref{tab:1} -- \ref{tab:5}
  following Table~\ref{tab:5}. 
\end{enumerate}
}
\end{table}
\begin{table}
\caption{
  Character tables and defining relations of the single-valued irreducible
  corepresentations of the magnetic group of the tetragonal
  antiferromagnetic structures (a) -- (d) in Fig.~\ref{fig:strukt} with
  $\alpha = \beta = 90^0$ (i.e., with the magnetic moments lying in
  $\pm z$ direction).  
\label{tab:5}
}
\begin{tabular}[t]{lcccccccccccc}
\multicolumn{13}{c}{$\Gamma (000)$ and $Z (00\frac{1}{2})$}\\
$\Gamma, Z:$ &  &  & $$ & $$ & $\{C^-_{4z}|000\}$ & $\{C_{2a}|\frac{1}{2}\frac{1}{2}0\}$ & $\{C_{2y}|\frac{1}{2}\frac{1}{2}0\}$ & $$ & $$ & $\{S^+_{4z}|000\}$ & $\{\sigma_{da}|\frac{1}{2}\frac{1}{2}0\}$ & $\{\sigma_y|\frac{1}{2}\frac{1}{2}0\}$\\
$\Gamma, Z:$ & $K$ & $\{K|\frac{1}{2}\frac{1}{2}0\}$ & $\{E|000\}$ & $\{C_{2z}|000\}$ & $\{C^+_{4z}|000\}$ & $\{C_{2b}|\frac{1}{2}\frac{1}{2}0\}$ & $\{C_{2x}|\frac{1}{2}\frac{1}{2}0\}$ & $\{I|000\}$ & $\{\sigma_z|000\}$ & $\{S^-_{4z}|000\}$ & $\{\sigma_{db}|\frac{1}{2}\frac{1}{2}0\}$ & $\{\sigma_x|\frac{1}{2}\frac{1}{2}0\}$\\
\hline
$\Gamma^+_1, Z^+_1$ & (a) & (a) & 1 & 1 & 1 & 1 & 1 & 1 & 1 & 1 & 1 & 1\\
$\Gamma^+_2, Z^+_2$ & (a) & (a) & 1 & 1 & 1 & -1 & -1 & 1 & 1 & 1 & -1 & -1\\
$\Gamma^+_3, Z^+_3$ & (a) & (a) & 1 & 1 & -1 & 1 & -1 & 1 & 1 & -1 & 1 & -1\\
$\Gamma^+_4, Z^+_4$ & (a) & (a) & 1 & 1 & -1 & -1 & 1 & 1 & 1 & -1 & -1 & 1\\
$\Gamma^+_5, Z^+_5$ & (a) & (a) & 2 & -2 & 0 & 0 & 0 & 2 & -2 & 0 & 0 & 0\\
$\Gamma^-_1, Z^-_1$ & (a) & (a) & 1 & 1 & 1 & 1 & 1 & -1 & -1 & -1 & -1 & -1\\
$\Gamma^-_2, Z^-_2$ & (a) & (a) & 1 & 1 & 1 & -1 & -1 & -1 & -1 & -1 & 1 & 1\\
$\Gamma^-_3, Z^-_3$ & (a) & (a) & 1 & 1 & -1 & 1 & -1 & -1 & -1 & 1 & -1 & 1\\
$\Gamma^-_4, Z^-_4$ & (a) & (a) & 1 & 1 & -1 & -1 & 1 & -1 & -1 & 1 & 1 & -1\\
$\Gamma^-_5, Z^-_5$ & (a) & (a) & 2 & -2 & 0 & 0 & 0 & -2 & 2 & 0 & 0 & 0\\
\hline\\
\end{tabular}\hspace{1cm}
\begin{tabular}[t]{lccccccccccc}
\multicolumn{12}{c}{$M (\frac{1}{2}\frac{1}{2}0)$ and $A
 (\frac{1}{2}\frac{1}{2}\frac{1}{2})$}\\ 
$M:$ &  &  & $$ & $$ & $$ & $$ & $\{\sigma_y|\frac{1}{2}\frac{3}{2}0\}$ & $\{\sigma_y|\frac{1}{2}\frac{1}{2}0\}$ & $\{C_{2a}|\frac{1}{2}\frac{3}{2}0\}$ & $\{C_{2a}|\frac{1}{2}\frac{1}{2}0\}$ & $\{S^-_{4z}|000\}$\\
$M:$ & $K$ & $\{K|\frac{1}{2}\frac{1}{2}0\}$ & $\{E|000\}$ & $\{C_{2z}|010\}$ & $\{E|010\}$ & $\{C_{2z}|000\}$ & $\{\sigma_x|\frac{1}{2}\frac{1}{2}0\}$ & $\{\sigma_x|\frac{1}{2}\frac{3}{2}0\}$ & $\{C_{2b}|\frac{1}{2}\frac{1}{2}0\}$ & $\{C_{2b}|\frac{1}{2}\frac{3}{2}0\}$ & $\{S^+_{4z}|010\}$\\
\hline
$A:$ &  &  & $$ & $$ & $$ & $$ & $\{\sigma_y|\frac{1}{2}\frac{1}{2}1\}$ & $\{\sigma_y|\frac{1}{2}\frac{1}{2}0\}$ & $\{C_{2a}|\frac{1}{2}\frac{1}{2}1\}$ & $\{C_{2a}|\frac{1}{2}\frac{1}{2}0\}$ & $\{S^-_{4z}|000\}$\\
$A:$ & $K$ & $\{K|\frac{1}{2}\frac{1}{2}0\}$ & $\{E|000\}$ & $\{C_{2z}|001\}$ & $\{E|001\}$ & $\{C_{2z}|000\}$ & $\{\sigma_x|\frac{1}{2}\frac{1}{2}0\}$ & $\{\sigma_x|\frac{1}{2}\frac{1}{2}1\}$ & $\{C_{2b}|\frac{1}{2}\frac{1}{2}0\}$ & $\{C_{2b}|\frac{1}{2}\frac{1}{2}1\}$ & $\{S^+_{4z}|001\}$\\
\hline
$M^+_1, A^+_1$ & (c) & (a) & 1 & 1 & -1 & -1 & 1 & -1 & i & -i & i\\
$M^+_2, A^+_2$ & (c) & (a) & 1 & 1 & -1 & -1 & 1 & -1 & -i & i & -i\\
$M^+_3, A^+_3$ & (c) & (a) & 1 & 1 & -1 & -1 & -1 & 1 & -i & i & i\\
$M^+_4, A^+_4$ & (c) & (a) & 1 & 1 & -1 & -1 & -1 & 1 & i & -i & -i\\
$M^+_5, A^+_5$ & (a) & (a) & 2 & -2 & -2 & 2 & 0 & 0 & 0 & 0 & 0\\
$M^-_1, A^-_1$ & (c) & (a) & 1 & 1 & -1 & -1 & 1 & -1 & i & -i & i\\
$M^-_2, A^-_2$ & (c) & (a) & 1 & 1 & -1 & -1 & 1 & -1 & -i & i & -i\\
$M^-_3, A^-_3$ & (c) & (a) & 1 & 1 & -1 & -1 & -1 & 1 & -i & i & i\\
$M^-_4, A^-_4$ & (c) & (a) & 1 & 1 & -1 & -1 & -1 & 1 & i & -i & -i\\
$M^-_5, A^-_5$ & (a) & (a) & 2 & -2 & -2 & 2 & 0 & 0 & 0 & 0 & 0\\
\hline\\
\end{tabular}\hspace{1cm}
\begin{tabular}[t]{lccccccccccc}
\multicolumn{12}{c}{$M (\frac{1}{2}\frac{1}{2}0)$ and $A (\frac{1}{2}\frac{1}{2}\frac{1}{2})$\qquad $(continued)$}\\
$M:$ & $\{S^-_{4z}|010\}$ & $$ & $$ & $$ & $$ & $\{C_{2y}|\frac{1}{2}\frac{1}{2}0\}$ & $\{C_{2y}|\frac{1}{2}\frac{3}{2}0\}$ & $\{\sigma_{da}|\frac{1}{2}\frac{1}{2}0\}$ & $\{\sigma_{da}|\frac{1}{2}\frac{3}{2}0\}$ & $\{C^+_{4z}|010\}$ & $\{C^+_{4z}|000\}$\\
$M:$ & $\{S^+_{4z}|000\}$ & $\{I|010\}$ & $\{\sigma_z|000\}$ & $\{I|000\}$ & $\{\sigma_z|010\}$ & $\{C_{2x}|\frac{1}{2}\frac{3}{2}0\}$ & $\{C_{2x}|\frac{1}{2}\frac{1}{2}0\}$ & $\{\sigma_{db}|\frac{1}{2}\frac{3}{2}0\}$ & $\{\sigma_{db}|\frac{1}{2}\frac{1}{2}0\}$ & $\{C^-_{4z}|000\}$ & $\{C^-_{4z}|010\}$\\
\hline
$A:$ & $\{S^-_{4z}|001\}$ & $$ & $$ & $$ & $$ & $\{C_{2y}|\frac{1}{2}\frac{1}{2}0\}$ & $\{C_{2y}|\frac{1}{2}\frac{1}{2}1\}$ & $\{\sigma_{da}|\frac{1}{2}\frac{1}{2}0\}$ & $\{\sigma_{da}|\frac{1}{2}\frac{1}{2}1\}$ & $\{C^+_{4z}|001\}$ & $\{C^+_{4z}|000\}$\\
$A:$ & $\{S^+_{4z}|000\}$ & $\{I|001\}$ & $\{\sigma_z|000\}$ & $\{I|000\}$ & $\{\sigma_z|001\}$ & $\{C_{2x}|\frac{1}{2}\frac{1}{2}1\}$ & $\{C_{2x}|\frac{1}{2}\frac{1}{2}0\}$ & $\{\sigma_{db}|\frac{1}{2}\frac{1}{2}1\}$ & $\{\sigma_{db}|\frac{1}{2}\frac{1}{2}0\}$ & $\{C^-_{4z}|000\}$ & $\{C^-_{4z}|001\}$\\
\hline
$M^+_1, A^+_1$ & -i & 1 & 1 & -1 & -1 & 1 & -1 & i & -i & i & -i\\
$M^+_2, A^+_2$ & i & 1 & 1 & -1 & -1 & 1 & -1 & -i & i & -i & i\\
$M^+_3, A^+_3$ & -i & 1 & 1 & -1 & -1 & -1 & 1 & -i & i & i & -i\\
$M^+_4, A^+_4$ & i & 1 & 1 & -1 & -1 & -1 & 1 & i & -i & -i & i\\
$M^+_5, A^+_5$ & 0 & 2 & -2 & -2 & 2 & 0 & 0 & 0 & 0 & 0 & 0\\
$M^-_1, A^-_1$ & -i & -1 & -1 & 1 & 1 & -1 & 1 & -i & i & -i & i\\
$M^-_2, A^-_2$ & i & -1 & -1 & 1 & 1 & -1 & 1 & i & -i & i & -i\\
$M^-_3, A^-_3$ & -i & -1 & -1 & 1 & 1 & 1 & -1 & i & -i & -i & i\\
$M^-_4, A^-_4$ & i & -1 & -1 & 1 & 1 & 1 & -1 & -i & i & i & -i\\
$M^-_5, A^-_5$ & 0 & -2 & 2 & 2 & -2 & 0 & 0 & 0 & 0 & 0 & 0\\
\hline\\
\end{tabular}\hspace{1cm}
\begin{tabular}[t]{lcccccccccccc}
\multicolumn{13}{c}{$R (0\frac{1}{2}\frac{1}{2})$ and $X
 (0\frac{1}{2}0)$}\\ 
$R:$ &  &  & $$ & $$ & $\{\sigma_{db}|\frac{1}{2}\frac{1}{2}1\}$ & $\{\sigma_{da}|\frac{1}{2}\frac{1}{2}1\}$ & $\{C_{2z}|000\}$ & $$ & $$ & $\{C_{2a}|\frac{1}{2}\frac{1}{2}1\}$ & $\{C_{2b}|\frac{1}{2}\frac{1}{2}1\}$ & $\{I|000\}$\\
$R:$ & $K$ & $\{K|\frac{1}{2}\frac{1}{2}0\}$ & $\{E|000\}$ & $\{E|001\}$ & $\{\sigma_{db}|\frac{1}{2}\frac{1}{2}0\}$ & $\{\sigma_{da}|\frac{1}{2}\frac{1}{2}0\}$ & $\{C_{2z}|001\}$ & $\{\sigma_z|000\}$ & $\{\sigma_z|001\}$ & $\{C_{2a}|\frac{1}{2}\frac{1}{2}0\}$ & $\{C_{2b}|\frac{1}{2}\frac{1}{2}0\}$ & $\{I|001\}$\\
\hline
$X:$ &  &  & $$ & $$ & $\{\sigma_{db}|\frac{1}{2}\frac{3}{2}0\}$ & $\{\sigma_{da}|\frac{1}{2}\frac{3}{2}0\}$ & $\{C_{2z}|000\}$ & $$ & $$ & $\{C_{2a}|\frac{1}{2}\frac{3}{2}0\}$ & $\{C_{2b}|\frac{1}{2}\frac{3}{2}0\}$ & $\{I|000\}$\\
$X:$ & $K$ & $\{K|\frac{1}{2}\frac{1}{2}0\}$ & $\{E|000\}$ & $\{E|010\}$ & $\{\sigma_{db}|\frac{1}{2}\frac{1}{2}0\}$ & $\{\sigma_{da}|\frac{1}{2}\frac{1}{2}0\}$ & $\{C_{2z}|010\}$ & $\{\sigma_z|000\}$ & $\{\sigma_z|010\}$ & $\{C_{2a}|\frac{1}{2}\frac{1}{2}0\}$ & $\{C_{2b}|\frac{1}{2}\frac{1}{2}0\}$ & $\{I|010\}$\\
\hline
$R_1, X_1$ & (a) & (a) & 2 & -2 & 0 & 0 & 0 & 2 & -2 & 0 & 0 & 0\\
$R_2, X_2$ & (a) & (a) & 2 & -2 & 0 & 0 & 0 & -2 & 2 & 0 & 0 & 0\\
\hline\\
\end{tabular}\hspace{1cm}
\end{table}
\begin{table}
\addtocounter{table}{-1}
\caption{continued}
\footnotetext{
\ \\
Notes to Table~\ref{tab:5}
\begin{enumerate}
\item The antiferromagnetic structures (a) -- (d) in Fig.~\ref{fig:strukt}
  with $\alpha = \beta = 90^0$ (i.e., with the magnetic moments lying
  in $\pm z$ direction) possess the space group $H = P4/mbm =
  \Gamma_qD^{5}_{4h}$ (127).
\item $K$ stands for the operator of time-inversion. $M = D^{5}_{4h} +
  \{K|\frac{1}{2}\frac{1}{2}0\}D^{5}_{4h}$ is the magnetic group of the
  antiferromagnetic structure.
\item The upper rows list the two anti-unitary elements $K$ and
  $\{K|\frac{1}{2}\frac{1}{2}0\}$ and the space group elements of the
  little group of the point(s) of symmetry given at the beginning of the
  row.
\item 
The coordinate system is given in
Fig.~\ref{fig:strukt} (c). However, the translations $\vec t_1, \vec t_2,
\vec t_3$ in the tetragonal lattice differ slightly from the given
translations $\vec T_1, \vec T_2, \vec T_3$ in the orthorhombic lattice: 
$$\vec t_1 = \vec T_2,\qquad \vec t_2 = - \vec T_1,\qquad \vec t_3 = \vec
T_3.$$ 
\item The antiferromagnetic state, if it exists, belongs to a
  corepresentation of $M$ which is derived from one of the representations
  at points $M$ or $A$ because the little groups of $M$ and $A$ comprise
  the whole space group and have one-dimensional representations following
  case $(c)$ and case $(a)$ with respect to $K$ and
  $\{K|\frac{1}{2}\frac{1}{2}0\}$, respectively.
\end{enumerate}
\ \\
Common notes to Tables~\ref{tab:1} -- \ref{tab:5}
\begin{enumerate}
\addtocounter{enumi}{5}
\item The tables are determined from Table 5.7 in the textbook of Bradley
  and Cracknell \protect\cite{bc}.  
\item  
The representations are labeled in the notation of Table 5.8 of
Bradley and Cracknell \protect\cite{bc}.
\item The labels of the point group operations are related to the $xyz$
  coordinates in Fig.~\ref{fig:strukt} and the translational parts of the
  space group elements are written in terms of the basic translations
  $T_1$, $T_2$, and $T_3$ given in Fig.~\ref{fig:strukt}, too.
\item 
  The rows following the names of representations (for example $\Gamma^+_1,
  Z_2$) list the characters of these representations.
\item As usual, space group elements and characters in the same column
  belong to the same class. The tables list all the elements of a
  class.  
\item The points of symmetry have the usual coordinates (given in the
  caption of the tables) in the Brillouin zone for $\Gamma_q$, $\Gamma_o$,
  and $\Gamma_m$, respectively, as given, e.g., in Table 3.6 of Bradley and
  Cracknell \protect\cite{bc}.
\item The elements $K$ and $\{K|\frac{1}{2}\frac{1}{2}0\}$ define two
  magnetic groups $M_1 = H + KH \quad\mbox{ and} \quad M_2 = H
  +\{K|\frac{1}{2}\frac{1}{2}0\}H$ (with $H$ being the space group). $M_2$
  is the group of the considered magnetic structure. The entries below $K$
  and $\{K|\frac{1}{2}\frac{1}{2}0\}$ specify whether the corepresentations
  of $M_1$ and $M_2$, respectively, derived from the representation(s)
  $R_i$ in the considered row follow case $(a)$, case $(b)$, or case $(c)$
  when they are given by Eqs.~(7.3.45-47) of Ref.~\protect\cite{bc}. In
  case $(a)$, there is no change in the degeneracy of $R_i$, and in case
  $(c)$ the degeneracy becomes doubled in the corepresentation of $M_1$ and
  $M_2$, respectively. In same cases, this information is given in separate
  tables underneath the character tables.
\item The cases $(a)$, $(b)$, and $(c)$ are determined by Eq.~(7.3.51) of
  Ref.~\protect\cite{bc}. 
\end{enumerate}
}
\end{table}
\begin{table}
\caption{
Compatibility relations between the Brillouin zone for tetragonal
paramagnetic YBa$_2$Cu$_3$O$_6$ and the Brillouin zone for the two
orthorhombic antiferromagnetic structures (a) and (b) depicted in
Fig.~\ref{fig:strukt}.  
\label{tab:6}
}
\begin{tabular}[t]{ccccccccccc}
\multicolumn{11}{c}{$\Gamma$}\\
\hline
&$\Gamma^+_1$ & $\Gamma^+_2$ & $\Gamma^+_3$ & $\Gamma^+_4$ & $\Gamma^+_5$ & 
$\Gamma^-_1$ & $\Gamma^-_2$ & $\Gamma^-_3$ & $\Gamma^-_4$ & $\Gamma^-_5$\\
Fm, Af:&$\Gamma^+_1$ & $\Gamma^+_2$ & $\Gamma^+_2$ & $\Gamma^+_1$ & $\Gamma^+_3$
 + $\Gamma^+_4$ & $\Gamma^-_1$ & $\Gamma^-_2$ & $\Gamma^-_2$ & $\Gamma^-_1$ & 
$\Gamma^-_3$ + $\Gamma^-_4$\\
\hline\\
\end{tabular}\hspace{1cm}
\begin{tabular}[t]{ccccccccccc}
\multicolumn{11}{c}{$\Gamma$}\\
\hline
&$M^+_1$ & $M^+_2$ & $M^+_3$ & $M^+_4$ & $M^+_5$ & $M^-_1$ & $M^-_2$ & 
$M^-_3$ & $M^-_4$ & $M^-_5$\\
Fm, Af:&$\Gamma^+_2$ & $\Gamma^+_1$ & $\Gamma^+_1$ & $\Gamma^+_2$ & $\Gamma^+_3$
 + $\Gamma^+_4$ & $\Gamma^-_2$ & $\Gamma^-_1$ & $\Gamma^-_1$ & $\Gamma^-_2$ & 
$\Gamma^-_3$ + $\Gamma^-_4$\\
\hline\\
\end{tabular}\hspace{1cm}
\begin{tabular}[t]{ccccccccccc}
\multicolumn{11}{c}{$\Gamma$}\\
\hline
&$Z^+_1$ & $Z^+_2$ & $Z^+_3$ & $Z^+_4$ & $Z^+_5$ & $Z^-_1$ & $Z^-_2$ & 
$Z^-_3$ & $Z^-_4$ & $Z^-_5$\\
Fm:&$\Gamma^+_3$ & $\Gamma^+_4$ & $\Gamma^+_4$ & $\Gamma^+_3$ & $\Gamma^+_1$
 + $\Gamma^+_2$ & $\Gamma^-_3$ & $\Gamma^-_4$ & $\Gamma^-_4$ & $\Gamma^-_3$ & 
$\Gamma^-_1$ + $\Gamma^-_2$\\
Af:&$\Gamma^-_4$ & $\Gamma^-_3$ & $\Gamma^-_3$ & $\Gamma^-_4$ & $\Gamma^-_1$
 + $\Gamma^-_2$ & $\Gamma^+_4$ & $\Gamma^+_3$ & $\Gamma^+_3$ & $\Gamma^+_4$ & 
$\Gamma^+_1$ + $\Gamma^+_2$\\
\hline\\
\end{tabular}\hspace{1cm}
\begin{tabular}[t]{ccccccccccc}
\multicolumn{11}{c}{$\Gamma$}\\
\hline
&$A^+_1$ & $A^+_2$ & $A^+_3$ & $A^+_4$ & $A^+_5$ & $A^-_1$ & $A^-_2$ & 
$A^-_3$ & $A^-_4$ & $A^-_5$\\
Fm:&$\Gamma^+_4$ & $\Gamma^+_3$ & $\Gamma^+_3$ & $\Gamma^+_4$ & $\Gamma^+_1$
 + $\Gamma^+_2$ & $\Gamma^-_4$ & $\Gamma^-_3$ & $\Gamma^-_3$ & $\Gamma^-_4$ & 
$\Gamma^-_1$ + $\Gamma^-_2$\\
Af:&$\Gamma^-_3$ & $\Gamma^-_4$ & $\Gamma^-_4$ & $\Gamma^-_3$ & $\Gamma^-_1$
 + $\Gamma^-_2$ & $\Gamma^+_3$ & $\Gamma^+_4$ & $\Gamma^+_4$ & $\Gamma^+_3$ & 
$\Gamma^+_1$ + $\Gamma^+_2$\\
\hline\\
\end{tabular}\hspace{1cm}
\begin{tabular}[t]{ccccccccc}
\multicolumn{9}{c}{$S$}\\
\hline
&$R^+_1$ & $R^+_2$ & $R^+_3$ & $R^+_4$ & $R^-_1$ & $R^-_2$ & $R^-_3$ & 
$R^-_4$\\
Fm:&$S^-_3$ + $S^-_4$ & $S^-_1$ + $S^-_2$ & $S^-_3$ + $S^-_4$ & $S^-_1$ + $S^-_2$ & 
$S^+_3$ + $S^+_4$ & $S^+_1$ + $S^+_2$ & $S^+_3$ + $S^+_4$ & $S^+_1$ + $S^+_2$\\
Af:&$S^-_1$ + $S^-_2$ & $S^-_3$ + $S^-_4$ & $S^-_1$ + $S^-_2$ & $S^-_3$ + $S^-_4$ & 
$S^+_1$ + $S^+_2$ & $S^+_3$ + $S^+_4$ & $S^+_1$ + $S^+_2$ & $S^+_3$ + $S^+_4$\\
\hline\\
\end{tabular}\hspace{1cm}
\begin{tabular}[t]{ccccccccc}
\multicolumn{9}{c}{$S$}\\
\hline
&$X^+_1$ & $X^+_2$ & $X^+_3$ & $X^+_4$ & $X^-_1$ & $X^-_2$ & $X^-_3$ & 
$X^-_4$\\
Fm:&$S^-_1$ + $S^-_2$ & $S^-_3$ + $S^-_4$ & $S^-_1$ + $S^-_2$ & $S^-_3$ + $S^-_4$ & 
$S^+_1$ + $S^+_2$ & $S^+_3$ + $S^+_4$ & $S^+_1$ + $S^+_2$ & $S^+_3$ + $S^+_4$\\
Af:&$S^+_3$ + $S^+_4$ & $S^+_1$ + $S^+_2$ & $S^+_3$ + $S^+_4$ & $S^+_1$ + $S^+_2$ & 
$S^-_3$ + $S^-_4$ & $S^-_1$ + $S^-_2$ & $S^-_3$ + $S^-_4$ & $S^-_1$ + $S^-_2$\\
\hline\\
\end{tabular}\hspace{1cm}
\footnotetext{
\ \\
Notes to Table~\ref{tab:6}
\begin{enumerate}
\item Both the fm-like structure (a) and the af-like structure (b) in
  Fig.~\ref{fig:strukt} possess the space group $\Gamma_oD^{16}_{2h}$ (62)
  with, however, different positions of the origin. Hence, the
  compatibility relations may differ.
\item The upper rows list the representations of the little groups of the
  points of symmetry in the Brillouin zone for the tetragonal paramagnetic
  phase (with the space group $\Gamma_qD^{1}_{4h}$). The rows following the
  label Fm or Af list representations of the little groups of the related
  points of symmetry in the Brillouin zone for the fm-like and the
  af-like antiferromagnetic structure, respectively.
\item The Brillouin zone for the fm-like and the af-like structure
  (with the Bravais lattice $\Gamma_o$) lies within the Brillouin zone for
  the tetragonal paramagnetic phase (with the Bravais lattice $\Gamma_q$),
  see Fig.~\ref{fig:BZ} (a).
\item See also the common notes to Tables~\ref{tab:6} -- \ref{tab:9}
  following Table~\ref{tab:9}.    
\end{enumerate}
}
\end{table}
\begin{table}
\caption{
Compatibility relations between the Brillouin zone for tetragonal
paramagnetic YBa$_2$Cu$_3$O$_6$ and the Brillouin zone for the orthorhombic
antiferromagnetic structure (d) depicted in Fig.~\ref{fig:strukt}.  
\label{tab:7}
}
\begin{tabular}[t]{cccccccccc}
\multicolumn{10}{c}{$\Gamma$}\\
\hline
$\Gamma^+_1$ & $\Gamma^+_2$ & $\Gamma^+_3$ & $\Gamma^+_4$ & $\Gamma^+_5$ & 
$\Gamma^-_1$ & $\Gamma^-_2$ & $\Gamma^-_3$ & $\Gamma^-_4$ & $\Gamma^-_5$\\
$\Gamma_1$ & $\Gamma_4$ & $\Gamma_4$ & $\Gamma_1$ & $\Gamma_3$ + $\Gamma_2$ & 
$\Gamma_3$ & $\Gamma_2$ & $\Gamma_2$ & $\Gamma_3$ & $\Gamma_1$ + $\Gamma_4$\\
\hline\\
\end{tabular}\hspace{1cm}
\begin{tabular}[t]{cccccccccc}
\multicolumn{10}{c}{$\Gamma$}\\
\hline
$M^+_1$ & $M^+_2$ & $M^+_3$ & $M^+_4$ & $M^+_5$ & $M^-_1$ & $M^-_2$ & 
$M^-_3$ & $M^-_4$ & $M^-_5$\\
$\Gamma_4$ & $\Gamma_1$ & $\Gamma_1$ & $\Gamma_4$ & $\Gamma_3$ + $\Gamma_2$ & 
$\Gamma_2$ & $\Gamma_3$ & $\Gamma_3$ & $\Gamma_2$ & $\Gamma_1$ + $\Gamma_4$\\
\hline\\
\end{tabular}\hspace{1cm}
\begin{tabular}[t]{cccccccccc}
\multicolumn{10}{c}{$Z$}\\
\hline
$Z^+_1$ & $Z^+_2$ & $Z^+_3$ & $Z^+_4$ & $Z^+_5$ & $Z^-_1$ & $Z^-_2$ & 
$Z^-_3$ & $Z^-_4$ & $Z^-_5$\\
$Z_1$ & $Z_4$ & $Z_4$ & $Z_1$ & $Z_3$ + $Z_2$ & $Z_3$ & $Z_2$ & $Z_2$ & 
$Z_3$ & $Z_1$ + $Z_4$\\
\hline\\
\end{tabular}\hspace{1cm}
\begin{tabular}[t]{cccccccccc}
\multicolumn{10}{c}{$Z$}\\
\hline
$A^+_1$ & $A^+_2$ & $A^+_3$ & $A^+_4$ & $A^+_5$ & $A^-_1$ & $A^-_2$ & 
$A^-_3$ & $A^-_4$ & $A^-_5$\\
$Z_4$ & $Z_1$ & $Z_1$ & $Z_4$ & $Z_3$ + $Z_2$ & $Z_2$ & $Z_3$ & $Z_3$ & 
$Z_2$ & $Z_1$ + $Z_4$\\
\hline\\
\end{tabular}\hspace{1cm}
\begin{tabular}[t]{cccccccc}
\multicolumn{8}{c}{$R$}\\
\hline
$R^+_1$ & $R^+_2$ & $R^+_3$ & $R^+_4$ & $R^-_1$ & $R^-_2$ & $R^-_3$ & 
$R^-_4$\\
$R_1$ + $R_2$ & $R_3$ + $R_4$ & $R_1$ + $R_2$ & $R_3$ + $R_4$ & $R_3$ + $R_4$ & 
$R_1$ + $R_2$ & $R_3$ + $R_4$ & $R_1$ + $R_2$\\
\hline\\
\end{tabular}\hspace{1cm}
\begin{tabular}[t]{cccccccc}
\multicolumn{8}{c}{$S$}\\
\hline
$X^+_1$ & $X^+_2$ & $X^+_3$ & $X^+_4$ & $X^-_1$ & $X^-_2$ & $X^-_3$ & 
$X^-_4$\\
$S_1$ + $S_2$ & $S_3$ + $S_4$ & $S_1$ + $S_2$ & $S_3$ + $S_4$ & $S_3$ + $S_4$ & 
$S_1$ + $S_2$ & $S_3$ + $S_4$ & $S_1$ + $S_2$\\
\hline\\
\end{tabular}\hspace{1cm}
\footnotetext{
\ \\
Notes to Table~\ref{tab:7}
\begin{enumerate}
\item The antiferromagnetic structure (d) in Fig.~\ref{fig:strukt} has the
  space group $\Gamma_oC^{2}_{2v}$ (26).
\item The upper rows list the representations of the little groups of the
  points of symmetry in the Brillouin zone for the tetragonal paramagnetic
  phase (with the space group $\Gamma_qD^{1}_{4h}$). The lower rows list
  representations of the little groups of the related points of symmetry in
  the Brillouin zone for the antiferromagnetic structure.
\item The Brillouin zone for the antiferromagnetic structure (with the
  Bravais lattice $\Gamma_o$) lies within the Brillouin zone for the
  tetragonal paramagnetic phase (with the Bravais lattice $\Gamma_q$), see
  Fig.~\ref{fig:BZ} (b).
\item See also the common notes to Tables~\ref{tab:6} -- \ref{tab:9}
  following Table~\ref{tab:9}.    
\end{enumerate}
}
\end{table}
\begin{table}
\caption{
Compatibility relations between the Brillouin zone for tetragonal
paramagnetic YBa$_2$Cu$_3$O$_6$ and the Brillouin zone for the monoclinic
antiferromagnetic structure (c) depicted in Fig.~\ref{fig:strukt}.  
\label{tab:8}
}
\begin{tabular}[t]{cccccccccc}
\multicolumn{10}{c}{$\Gamma$}\\
\hline
$\Gamma^+_1$ & $\Gamma^+_2$ & $\Gamma^+_3$ & $\Gamma^+_4$ & $\Gamma^+_5$ & 
$\Gamma^-_1$ & $\Gamma^-_2$ & $\Gamma^-_3$ & $\Gamma^-_4$ & $\Gamma^-_5$\\
$\Gamma^+_1$ & $\Gamma^+_2$ & $\Gamma^+_2$ & $\Gamma^+_1$ & $\Gamma^+_1$
 + $\Gamma^+_2$ & $\Gamma^-_1$ & $\Gamma^-_2$ & $\Gamma^-_2$ & $\Gamma^-_1$ & 
$\Gamma^-_1$ + $\Gamma^-_2$\\
\hline\\
\end{tabular}\hspace{1cm}
\begin{tabular}[t]{cccccccccc}
\multicolumn{10}{c}{$\Gamma$}\\
\hline
$M^+_1$ & $M^+_2$ & $M^+_3$ & $M^+_4$ & $M^+_5$ & $M^-_1$ & $M^-_2$ & 
$M^-_3$ & $M^-_4$ & $M^-_5$\\
$\Gamma^+_2$ & $\Gamma^+_1$ & $\Gamma^+_1$ & $\Gamma^+_2$ & $\Gamma^+_1$
 + $\Gamma^+_2$ & $\Gamma^-_2$ & $\Gamma^-_1$ & $\Gamma^-_1$ & $\Gamma^-_2$ & 
$\Gamma^-_1$ + $\Gamma^-_2$\\
\hline\\
\end{tabular}\hspace{1cm}
\begin{tabular}[t]{cccccccccc}
\multicolumn{10}{c}{$Z$}\\
\hline
$Z^+_1$ & $Z^+_2$ & $Z^+_3$ & $Z^+_4$ & $Z^+_5$ & $Z^-_1$ & $Z^-_2$ & 
$Z^-_3$ & $Z^-_4$ & $Z^-_5$\\
$Z^+_1$ & $Z^+_2$ & $Z^+_2$ & $Z^+_1$ & $Z^+_1$ + $Z^+_2$ & $Z^-_1$ & 
$Z^-_2$ & $Z^-_2$ & $Z^-_1$ & $Z^-_1$ + $Z^-_2$\\
\hline\\
\end{tabular}\hspace{1cm}
\begin{tabular}[t]{cccccccccc}
\multicolumn{10}{c}{$Z$}\\
\hline
$A^+_1$ & $A^+_2$ & $A^+_3$ & $A^+_4$ & $A^+_5$ & $A^-_1$ & $A^-_2$ & 
$A^-_3$ & $A^-_4$ & $A^-_5$\\
$Z^+_2$ & $Z^+_1$ & $Z^+_1$ & $Z^+_2$ & $Z^+_1$ + $Z^+_2$ & $Z^-_2$ & 
$Z^-_1$ & $Z^-_1$ & $Z^-_2$ & $Z^-_1$ + $Z^-_2$\\
\hline\\
\end{tabular}\hspace{1cm}
\begin{tabular}[t]{cccccccc}
\multicolumn{8}{c}{$E$}\\
\hline
$R^+_1$ & $R^+_2$ & $R^+_3$ & $R^+_4$ & $R^-_1$ & $R^-_2$ & $R^-_3$ & 
$R^-_4$\\
$E^+_1$ + $E^+_2$ & $E^+_1$ + $E^+_2$ & $E^+_1$ + $E^+_2$ & $E^+_1$ + $E^+_2$ & 
$E^-_1$ + $E^-_2$ & $E^-_1$ + $E^-_2$ & $E^-_1$ + $E^-_2$ & $E^-_1$ + $E^-_2$\\
\hline\\
\end{tabular}\hspace{1cm}
\begin{tabular}[t]{cccccccc}
\multicolumn{8}{c}{$A$}\\
\hline
$X^+_1$ & $X^+_2$ & $X^+_3$ & $X^+_4$ & $X^-_1$ & $X^-_2$ & $X^-_3$ & 
$X^-_4$\\
$A^+_1$ + $A^+_2$ & $A^+_1$ + $A^+_2$ & $A^+_1$ + $A^+_2$ & $A^+_1$ + $A^+_2$ & 
$A^-_1$ + $A^-_2$ & $A^-_1$ + $A^-_2$ & $A^-_1$ + $A^-_2$ & $A^-_1$ + $A^-_2$\\
\hline\\
\end{tabular}\hspace{1cm}
\footnotetext{
\ \\
Notes to Table~\ref{tab:8}
\begin{enumerate}
\item The antiferromagnetic structure (c) in Fig.~\ref{fig:strukt} has the
  space group $\Gamma_mC^{5}_{2h}$ (14).
\item The upper rows list the representations of the little groups of the
  points of symmetry in the Brillouin zone for the tetragonal paramagnetic
  phase (with the space group $\Gamma_qD^{1}_{4h}$). The lower rows list
  representations of the little groups of the related points of symmetry in
  the Brillouin zone for the antiferromagnetic structure.
\item The Brillouin zone for the antiferromagnetic structure (with the
  Bravais lattice $\Gamma_m$) lies within the Brillouin zone for the
  tetragonal paramagnetic phase (with the Bravais lattice $\Gamma_q$). The
  Brillouin zone for $\Gamma_m$ is not drawn in this paper, but, e.g., in
  Fig.~3.3 of Ref.~\cite{bc}.
\item See also the common notes to Tables~\ref{tab:6} -- \ref{tab:9}
  following Table~\ref{tab:9}.    
\end{enumerate}
}
\end{table}
\begin{table}
\caption{
Compatibility relations between the Brillouin zone for tetragonal
paramagnetic YBa$_2$Cu$_3$O$_6$ and the Brillouin zone for the tetragonal
antiferromagnetic structure with $\alpha = \beta = 90^0$ in all the
structures of Fig.~\ref{fig:strukt}. 
\label{tab:9}
}
\begin{tabular}[t]{cccccccccc}
\hline\\
\end{tabular}\hspace{1cm}
\begin{tabular}[t]{cccccccccc}
\multicolumn{10}{c}{$\Gamma$}\\
\hline
$\Gamma^+_1$ & $\Gamma^+_2$ & $\Gamma^+_3$ & $\Gamma^+_4$ & $\Gamma^+_5$ & 
$\Gamma^-_1$ & $\Gamma^-_2$ & $\Gamma^-_3$ & $\Gamma^-_4$ & $\Gamma^-_5$\\
$\Gamma^+_1$ & $\Gamma^+_2$ & $\Gamma^+_4$ & $\Gamma^+_3$ & $\Gamma^+_5$ & 
$\Gamma^-_1$ & $\Gamma^-_2$ & $\Gamma^-_4$ & $\Gamma^-_3$ & $\Gamma^-_5$\\
\hline\\
\end{tabular}\hspace{1cm}
\begin{tabular}[t]{cccccccccc}
\multicolumn{10}{c}{$\Gamma$}\\
\hline
$M^+_1$ & $M^+_2$ & $M^+_3$ & $M^+_4$ & $M^+_5$ & $M^-_1$ & $M^-_2$ & 
$M^-_3$ & $M^-_4$ & $M^-_5$\\
$\Gamma^+_2$ & $\Gamma^+_1$ & $\Gamma^+_3$ & $\Gamma^+_4$ & $\Gamma^+_5$ & 
$\Gamma^-_2$ & $\Gamma^-_1$ & $\Gamma^-_3$ & $\Gamma^-_4$ & $\Gamma^-_5$\\
\hline\\
\end{tabular}\hspace{1cm}
\begin{tabular}[t]{cccccccccc}
\multicolumn{10}{c}{$Z$}\\
\hline
$Z^+_1$ & $Z^+_2$ & $Z^+_3$ & $Z^+_4$ & $Z^+_5$ & $Z^-_1$ & $Z^-_2$ & 
$Z^-_3$ & $Z^-_4$ & $Z^-_5$\\
$Z^+_1$ & $Z^+_2$ & $Z^+_4$ & $Z^+_3$ & $Z^+_5$ & $Z^-_1$ & $Z^-_2$ & 
$Z^-_4$ & $Z^-_3$ & $Z^-_5$\\
\hline\\
\end{tabular}\hspace{1cm}
\begin{tabular}[t]{cccccccccc}
\multicolumn{10}{c}{$Z$}\\
\hline
$A^+_1$ & $A^+_2$ & $A^+_3$ & $A^+_4$ & $A^+_5$ & $A^-_1$ & $A^-_2$ & 
$A^-_3$ & $A^-_4$ & $A^-_5$\\
$Z^+_2$ & $Z^+_1$ & $Z^+_3$ & $Z^+_4$ & $Z^+_5$ & $Z^-_2$ & $Z^-_1$ & 
$Z^-_3$ & $Z^-_4$ & $Z^-_5$\\
\hline\\
\end{tabular}\hspace{1cm}
\begin{tabular}[t]{cccccccc}
\multicolumn{8}{c}{$A$}\\
\hline
$R^+_1$ & $R^+_2$ & $R^+_3$ & $R^+_4$ & $R^-_1$ & $R^-_2$ & $R^-_3$ & 
$R^-_4$\\
$A^-_5$ & $A^-_3$ + $A^-_4$ & $A^-_5$ & $A^-_1$ + $A^-_2$ & $A^+_5$ & 
$A^+_1$ + $A^+_2$ & $A^+_5$ & $A^+_3$ + $A^+_4$\\
\hline\\
\end{tabular}
\begin{tabular}[t]{cccccccc}
\multicolumn{8}{c}{$M$}\\
\hline
$X^+_1$ & $X^+_2$ & $X^+_3$ & $X^+_4$ & $X^-_1$ & $X^-_2$ & $X^-_3$ & 
$X^-_4$\\
$M^-_5$ & $M^-_3$ + $M^-_4$ & $M^-_5$ & $M^-_1$ + $M^-_2$ & $M^+_5$ & 
$M^+_1$ + $M^+_2$ & $M^+_5$ & $M^+_3$ + $M^+_4$\\
\hline\\
\end{tabular}
\footnotetext{
\ \\
Notes to Table~\ref{tab:9}
\begin{enumerate}
\item The tetragonal antiferromagnetic structure with $\alpha = \beta =
  90^0$ in all the structures in Fig.~\ref{fig:strukt} (i.e., with the
  magnetic moments lying in $\pm z$ direction) possesses the space group
  $\Gamma_qD^{5}_{4h}$ (127).
\item The upper rows list the representations of the little groups of the
  points of symmetry in the Brillouin zone for the tetragonal paramagnetic
  phase (with the space group $\Gamma_qD^{1}_{4h}$). The lower rows list
  representations of the little groups of the related points of symmetry in
  the Brillouin zone for the antiferromagnetic structure.
\item Both the antiferromagnetic structure and the paramagnetic phase have
  the Brillouin zone for $\Gamma_q$. The Brillouin zone for the
  antiferromagnetic structure is smaller than the Brillouin zone for the
  paramagnetic phase and lies within this Brillouin zone.
\end{enumerate}
\ \\
Common notes to Tables~\ref{tab:6} -- \ref{tab:9}
\begin{enumerate}
\addtocounter{enumi}{3}
\item The compatibility relations are determined in the way described in
  great detail in Ref.~\cite{eabf}.
\item The representations are labeled as given in Tables~\ref{tab:1}
  -- \ref{tab:5}.
\item The representations in the same column are compatible in the
  following sense: Bloch functions that are basis functions of a
  representation $R_i$ in the upper row can be unitarily transformed into
  basis functions of the representation given below $R_i$.
\end{enumerate}
}
\end{table}
\begin{table}
\caption{
Single-valued representations of all the energy bands in tetragonal
paramagnetic YBa$_2$Cu$_3$O$_6$ with symmetry-adapted and optimally
localized Wannier functions centered at the Cu atoms. 
\label{tab:10}
}
\begin{tabular}[t]{lccccccc}
&& $\Gamma$ & $M$ & $Z$ & $A$ & $R$ & $X$\\
\hline
Band 1&\ \ & 2$\Gamma^+_1$ + $\Gamma^-_2$ & 
2$M^+_1$ + $M^-_2$ & 2$Z^+_1$ + $Z^-_2$ & 2$A^+_1$ + $A^-_2$ & 2$R^+_1$
 + $R^-_3$ & 2$X^+_1$ + $X^-_3$\\
Band 2&& 2$\Gamma^+_2$ + $\Gamma^-_1$ & 2$M^+_2$
 + $M^-_1$ & 2$Z^+_2$ + $Z^-_1$ & 2$A^+_2$ + $A^-_1$ & 2$R^+_3$ + $R^-_1$ & 
2$X^+_3$ + $X^-_1$\\
Band 3&& 2$\Gamma^+_3$ + $\Gamma^-_4$ & 2$M^+_3$
 + $M^-_4$ & 2$Z^+_3$ + $Z^-_4$ & 2$A^+_3$ + $A^-_4$ & 2$R^+_1$ + $R^-_3$ & 
2$X^+_1$ + $X^-_3$\\
Band 4&& 2$\Gamma^+_4$ + $\Gamma^-_3$ & 2$M^+_4$
 + $M^-_3$ & 2$Z^+_4$ + $Z^-_3$ & 2$A^+_4$ + $A^-_3$ & 2$R^+_3$ + $R^-_1$ & 
2$X^+_3$ + $X^-_1$\\
Band 5&& $\Gamma^+_2$ + 2$\Gamma^-_1$ & $M^+_2$
 + 2$M^-_1$ & $Z^+_2$ + 2$Z^-_1$ & $A^+_2$ + 2$A^-_1$ & $R^+_3$ + 2$R^-_1$ & 
$X^+_3$ + 2$X^-_1$\\
Band 6&& $\Gamma^+_1$ + 2$\Gamma^-_2$ & $M^+_1$
 + 2$M^-_2$ & $Z^+_1$ + 2$Z^-_2$ & $A^+_1$ + 2$A^-_2$ & $R^+_1$ + 2$R^-_3$ & 
$X^+_1$ + 2$X^-_3$\\
Band 7&& $\Gamma^+_4$ + 2$\Gamma^-_3$ & $M^+_4$
 + 2$M^-_3$ & $Z^+_4$ + 2$Z^-_3$ & $A^+_4$ + 2$A^-_3$ & $R^+_3$ + 2$R^-_1$ & 
$X^+_3$ + 2$X^-_1$\\
Band 8&& $\Gamma^+_3$ + 2$\Gamma^-_4$ & $M^+_3$
 + 2$M^-_4$ & $Z^+_3$ + 2$Z^-_4$ & $A^+_3$ + 2$A^-_4$ & $R^+_1$ + 2$R^-_3$ & 
$X^+_1$ + 2$X^-_3$\\
\hline\\
\end{tabular}\hspace{1cm}
\footnotetext{
\ \\
Notes to Table~\ref{tab:10}
\begin{enumerate}
\item Tetragonal paramagnetic YBa$_2$Cu$_3$O$_6$ has the space group
  $\Gamma_qD^{1}_{4h}$ (123).
\item Each row defines one band consisting of three branches, because there
  are three Cu atoms in the unit cell.
\item Only bands with exactly one Wannier function on
each Cu site are given. 
\item See also the common notes to Tables~\ref{tab:10} -- \ref{tab:15}
  following Table~\ref{tab:15}.    
\end{enumerate}
}
\end{table}
\begin{table}
\caption{
Single-valued representations of all the antiferromagnetic energy bands
in the orthorhombic fm-like antiferromagnetic 
structure (a) depicted in Fig.~\ref{fig:strukt} with
symmetry-adapted and optimally localized Wannier functions centered on
the Cu sites. 
\label{tab:11}
}
\begin{tabular}[t]{cccccc}
&\ \ & $\Gamma$ & $Z$ & $Y$ & $X$\\
\hline
Band 1 && 2$\Gamma^+_1$ + 2$\Gamma^+_2$ 
 + 2$\Gamma^+_3$ + 2$\Gamma^+_4$ + $\Gamma^-_1$ + $\Gamma^-_2$ +
 $\Gamma^-_3$ 
 + $\Gamma^-_4$ & 3$Z_1$ + 3$Z_2$ & 3$Y_1$ + 3$Y_2$ & 3$X_1$ + 3$X_2$  \\
Band 2&&
$\Gamma^+_1$ + $\Gamma^+_2$ + $\Gamma^+_3$ 
 + $\Gamma^+_4$ + 2$\Gamma^-_1$ + 2$\Gamma^-_2$ + 2$\Gamma^-_3$ +
 2$\Gamma^-_4$ &  
3$Z_1$ + 3$Z_2$ & 3$Y_1$ + 3$Y_2$ & 3$X_1$ + 3$X_2$ \\
\hline\\
\end{tabular}
\begin{tabular}[t]{cccccc}
\multicolumn{6}{c}{$(continued)$}\\
&\ \ & $S$ & $U$ & $T$ & $R$\\
\hline
Band 1 && 2$S^-_1$ + 2$S^-_2$ + 2$S^-_3$
 + 2$S^-_4$ + $S^+_1$ + $S^+_2$ + $S^+_3$ + $S^+_4$ & 3$U_1$ + 3$U_2$ & 
3$T_1$ + 3$T_2$ & 3$R_1$ + 3$R_2$\\
Band 2 && $S^-_1$ + $S^-_2$ + $S^-_3$ + $S^-_4$
 + 2$S^+_1$ + 2$S^+_2$ + 2$S^+_3$ + 2$S^+_4$ & 3$U_1$ + 3$U_2$ & 3$T_1$
 + 3$T_2$ & 3$R_1$ + 3$R_2$\\
\hline\\
\end{tabular}
\footnotetext{
\ \\
Notes to Table~\ref{tab:11}
\begin{enumerate}
\item The fm-like antiferromagnetic structure (a) has the space group
  $\Gamma_qD^{16}_{2h}$ (62). 
\item The af-like structure (b) possesses the same space group with,
  however, a different position of the coordinate system. Therefore, this
  table differs from the following Table~\ref{tab:12}.
\item Each row defines one band consisting of twelve branches, because
  there are twelve Cu atoms in the unit cell.
\item See also the common notes to Tables~\ref{tab:10} -- \ref{tab:15}
  following Table~\ref{tab:15}.    
\end{enumerate}
}
\end{table}
\begin{table}
\caption{
Single-valued representations of all the antiferromagnetic energy bands
in the orthorhombic af-like antiferromagnetic 
structure (b) depicted in Fig.~\ref{fig:strukt} with
symmetry-adapted and optimally localized Wannier functions centered on
the Cu sites. 
\label{tab:12}
}
\begin{tabular}[t]{cccccc}
&\ \ & $\Gamma$ & $Z$ & $Y$ & $X$\\
\hline
Band 1 &&
2$\Gamma^+_1$ + 2$\Gamma^+_2$ 
 + $\Gamma^+_3$ + $\Gamma^+_4$ + $\Gamma^-_1$ + $\Gamma^-_2$ +
 2$\Gamma^-_3$  
 + 2$\Gamma^-_4$ & 3$Z_1$ + 3$Z_2$ & 4$Y_1$ + 2$Y_2$ & 4$X_1$ + 2$X_2$ \\
Band 2 && $\Gamma^+_1$ + $\Gamma^+_2$ + 2$\Gamma^+_3$
 + 2$\Gamma^+_4$ + 2$\Gamma^-_1$ + 2$\Gamma^-_2$ + $\Gamma^-_3$ +
 $\Gamma^-_4$ &  3$Z_1$ + 3$Z_2$ & 2$Y_1$ + 4$Y_2$ & 2$X_1$ + 4$X_2$\\
\hline\\
\end{tabular}
\begin{tabular}[t]{cccccc}
\multicolumn{6}{c}{$(continued)$}\\
&\ \ & $S$ & $U$ & $T$ & $R$\\
\hline
Band 1 && 2$S^-_1$ + 2$S^-_2$
 + $S^-_3$ + $S^-_4$ + $S^+_1$ + $S^+_2$ + 2$S^+_3$ + 2$S^+_4$ & 3$U_1$
 + 3$U_2$ & 3$T_1$ + 3$T_2$ & 3$R_1$ + 3$R_2$\\
Band 2 && $S^-_1$ + $S^-_2$
 + 2$S^-_3$ + 2$S^-_4$ + 2$S^+_1$ + 2$S^+_2$ + $S^+_3$ + $S^+_4$ & 3$U_1$
 + 3$U_2$ & 3$T_1$ + 3$T_2$ & 3$R_1$ + 3$R_2$\\
\hline\\
\end{tabular}
\footnotetext{
\ \\
Notes to Table~\ref{tab:12}
\begin{enumerate}
\item The af-like antiferromagnetic structure has the space group
  $\Gamma_qD^{16}_{2h}$ (62). 
\item The fm-like structure possesses the same space group with, however, a
  different position of the coordinate system. Therefore, this table
  differs from the foregoing Table~\ref{tab:11}.
\item Each row defines one band consisting of twelve branches, because
  there are twelve Cu atoms in the unit cell.
\item See also the common notes to Tables~\ref{tab:10} -- \ref{tab:15}
  following Table~\ref{tab:15}.    
\end{enumerate}
}
\end{table}
\begin{table}
\caption{
Single-valued representations of all the antiferromagnetic energy bands
in the orthorhombic antiferromagnetic structure (d) depicted in
Fig.~\ref{fig:strukt} with 
symmetry-adapted and optimally localized Wannier functions centered on
the Cu sites. 
\label{tab:13}
}
\begin{tabular}[t]{cccccc}
&\ \ & $\Gamma$ & $X$ & $Z$ & $Y$\\
\hline
Band 1 && 2$\Gamma_1$ + $\Gamma_3$ + $\Gamma_2$
 + 2$\Gamma_4$ & 2$X_1$ + $X_3$ + $X_2$ + 2$X_4$ & 2$Z_1$ + $Z_3$ + $Z_2$
 + 2$Z_4$ & 2$Y_1$ + 2$Y_2$ + $Y_3$ + $Y_4$\\
Band 2 && $\Gamma_1$ + 2$\Gamma_3$ + 2$\Gamma_2$
 + $\Gamma_4$ & $X_1$ + 2$X_3$ + 2$X_2$ + $X_4$ & $Z_1$ + 2$Z_3$ + 2$Z_2$
 + $Z_4$ & $Y_1$ + $Y_2$ + 2$Y_3$ + 2$Y_4$\\
\hline\\
\end{tabular}
\begin{tabular}[t]{cccccc}
\multicolumn{6}{c}{$(continued)$}\\
&\ \ & $T$ & $S$ & $U$ & $R$\\
\hline
Band 1 && 2$T_1$ + 2$T_2$ + $T_3$ + $T_4$ & 
2$S_1$ + 2$S_2$ + $S_3$ + $S_4$ & 2$U_1$ + $U_3$ + $U_2$ + 2$U_4$ & 2$R_1$
 + 2$R_2$ + $R_3$ + $R_4$\\
Band 2 && $T_1$ + $T_2$ + 2$T_3$ + 2$T_4$ & $S_1$ + $S_2$
 + 2$S_3$ + 2$S_4$ & $U_1$ + 2$U_3$ + 2$U_2$ + $U_4$ & $R_1$ + $R_2$ + 2$R_3$
 + 2$R_4$\\
\hline\\
\end{tabular}
\footnotetext{
\ \\
Notes to Table~\ref{tab:13}
\begin{enumerate}
\item The antiferromagnetic structure (d) in Fig.~\ref{fig:strukt} has
  the space group $\Gamma_qC^{2}_{2v}$ (26).
\item Each row defines one band consisting of six branches, because there
  are six Cu atoms in the unit cell.
\item See also the common notes to Tables~\ref{tab:10} -- \ref{tab:15}
  following Table~\ref{tab:15}.    
\end{enumerate}
}
\end{table}
\begin{table}
\caption{
Single-valued representations of all the antiferromagnetic energy bands
in the monoclinic antiferromagnetic structure (c) depicted in
Fig.~\ref{fig:strukt} with symmetry-adapted and optimally localized Wannier
functions centered on the Cu sites. 
\label{tab:14}
}
\begin{tabular}[t]{ccccc}
&\ \ & $\Gamma$ & $B$ & $Z$\\
\hline
Band 1&& 2$\Gamma^+_1$ + $\Gamma^-_1$
 + 2$\Gamma^+_2$ + $\Gamma^-_2$ & 3$B_1$ & 2$Z^+_1$ + $Z^-_1$ + 2$Z^+_2$
 + $Z^-_2$\\
Band 2&& $\Gamma^+_1$ + 2$\Gamma^-_1$ +
$\Gamma^+_2$ 
 + 2$\Gamma^-_2$ & 3$B_1$ & $Z^+_1$ + 2$Z^-_1$ + $Z^+_2$ + 2$Z^-_2$\\
\hline\\
\end{tabular}
\begin{tabular}[t]{ccccccc}
\multicolumn{7}{c}{$(continued)$}\\
&\ \ & $Y$ & $C$ & $A$ & $D$ & $E$\\
\hline
Band 1 && 3$Y_1$ & 3$C_1$ & 2$A^+_1$ + 2$A^+_2$
 + $A^-_1$ + $A^-_2$ & 3$D_1$ & 2$E^+_1$ + 2$E^+_2$ + $E^-_1$ + $E^-_2$\\
Band 2 && 3$Y_1$ & 3$C_1$ & $A^+_1$ + $A^+_2$
 + 2$A^-_1$ + 2$A^-_2$ & 3$D_1$ & $E^+_1$ + $E^+_2$ + 2$E^-_1$ + 2$E^-_2$\\
\hline\\
\end{tabular}
\footnotetext{
\ \\
Notes to Table~\ref{tab:14}
\begin{enumerate}
\item The antiferromagnetic structure (c) in Fig.~\ref{fig:strukt} has the
  space group $\Gamma_mC^{5}_{2h}$ (14).
\item Each row defines one band consisting of six branches, because there
  are six Cu atoms in the unit cell.
\item See also the common notes to Tables~\ref{tab:10} -- \ref{tab:15}
  following Table~\ref{tab:15}.    
\end{enumerate}
}
\end{table}
\begin{table}
\caption{
Single-valued representations of all the antiferromagnetic energy bands
in the tetragonal antiferromagnetic structure with symmetry-adapted and 
optimally localized Wannier functions centered on the Cu sites.
\label{tab:15}
}
\begin{tabular}[t]{cccccccc}
&\ \ & $\Gamma$ & $M$ & $Z$ & $A$ & $R$ & $X$\\
\hline
Band 1 && 2$\Gamma^+_1$ + 2$\Gamma^+_2$
 + $\Gamma^-_1$ + $\Gamma^-_2$ & $M^+_5$ + 2$M^-_5$ & 2$Z^+_1$ + 2$Z^+_2$
 + $Z^-_1$ + $Z^-_2$ & $A^+_5$ + 2$A^-_5$ & 2$R_1$ + $R_2$ & 2$X_1$ + $X_2$\\
Band 2 && 2$\Gamma^+_3$ + 2$\Gamma^+_4$
 + $\Gamma^-_3$ + $\Gamma^-_4$ & $M^+_5$ + 2$M^-_5$ & 2$Z^+_3$ + 2$Z^+_4$
 + $Z^-_3$ + $Z^-_4$ & $A^+_5$ + 2$A^-_5$ & 2$R_1$ + $R_2$ & 2$X_1$ + $X_2$\\
Band 3 && $\Gamma^+_1$ + $\Gamma^+_2$ + 2$\Gamma^-_1$
 + 2$\Gamma^-_2$ & 2$M^+_5$ + $M^-_5$ & $Z^+_1$ + $Z^+_2$ + 2$Z^-_1$ + 2$Z^-_2$ & 
2$A^+_5$ + $A^-_5$ & $R_1$ + 2$R_2$ & $X_1$ + 2$X_2$\\
Band 4 && $\Gamma^+_3$ + $\Gamma^+_4$ + 2$\Gamma^-_3$
 + 2$\Gamma^-_4$ & 2$M^+_5$ + $M^-_5$ & $Z^+_3$ + $Z^+_4$ + 2$Z^-_3$ + 2$Z^-_4$ & 
2$A^+_5$ + $A^-_5$ & $R_1$ + 2$R_2$ & $X_1$ + 2$X_2$\\
\hline\\
\end{tabular}\hspace{1cm}
\footnotetext{
\ \\
Notes to Table~\ref{tab:15}
\begin{enumerate}
\item In the tetragonal antiferromagnetic structure the magnetic
  moments lie in $\pm z$ direction. It is obtained from the structures (a)
  -- (d) depicted Fig.~\ref{fig:strukt} by putting $\alpha = \beta = 90^0$.
\item This structure has the space group $\Gamma_qD^{5}_{4h}$ (127).
\item Each row defines one band consisting of six branches, because there
  are six Cu atoms in the unit cell.
\item Only bands with exactly one Wannier function on each Cu site are
  given.
\end{enumerate}
\ \\
Common notes to Tables~\ref{tab:10} -- \ref{tab:15}
\begin{enumerate}
\addtocounter{enumi}{3}
\item The bands are determined by Eq.~(B7) of Ref.~\protect\cite{la2cuo4}. 
\item The representations are labeled as given in Tables~\ref{tab:1}
  -- \ref{tab:5}.
\item Consider, for example, Table~\ref{tab:15}. Assume a band of the
  symmetry in any row of Table~\ref{tab:15} to exist in the band structure
  of a given metal and assume this band not to be connected to other bands.
  Then the Bloch functions of this band can be unitarily transformed into
  Wannier functions that are
\begin{itemize}
\item as well localized as possible; 
\item centered at the Cu atoms;
\item symmetry-adapted to the tetragonal antiferromagnetic structure with
  the magnetic group $D^{5}_{4h} +
  \{K|\frac{1}{2}\frac{1}{2}0\}D^{5}_{4h}$.\end{itemize}
\end{enumerate}
}
\end{table}

\end{document}